\documentclass[amsmath,amssymb,aps,nofootinbib,prd,preprint,tightenlines]{revtex4-2}

\usepackage{bm}
\usepackage[colorlinks,anchorcolor=black,citecolor=violet,linkcolor=blue]{hyperref}
\usepackage[mathscr]{euscript}
\usepackage{graphicx}
\usepackage{multirow,slashed}
\usepackage{xcolor}

\begin{document}
\baselineskip=17pt \parskip=5pt


\title{\boldmath$\Delta S$\,=\,2 nonleptonic hyperon decays as probes of new physics}

\author{Xiao-Gang He$,^{1,2}$ Jusak Tandean$,^3$ and German Valencia$^4$ \bigskip \\ \it
$^1$Tsung-Dao Lee Institute, KLPAC and SKLPPC Laboratories, School of Physics and Astronomy, \\
Shanghai Jiao Tong University, Shanghai 201210, China \medskip \\
$^2$Department of Physics, National Taiwan University, Taipei 10617, Taiwan \medskip \\
$^3$Tsung-Dao Lee Institute, Shanghai Jiao Tong University, Shanghai 201210, China \medskip \\
$^4$School of Physics and Astronomy, Monash University, Wellington Road, Clayton, VIC-3800, Australia \bigskip \\
\rm Abstract
\medskip \\
\begin{minipage}{0.93\textwidth} \baselineskip=17pt \parindent=3ex \small
Hyperon nonleptonic decays that change strangeness by two units, such as $\mathit\Xi\to N\pi$ and \mbox{$\mathit\Omega^-\to nK^-,\mathit\Lambda\pi^-,\mathit\Sigma^{(*)}\pi$},\, are highly suppressed in the standard model. Only a few of them have been searched for to date, leading to experimental upper bounds which are many orders of magnitude above the expectations of the standard model.
This leaves ample opportunity to look for indications of new physics in these processes. At the same time, most, but not all, $\Delta S$\,=\,2 interactions beyond the standard model are severely constrained by kaon-mixing data. We present two scenarios where new physics satisfying the kaon-mixing constraints can enhance the hyperon decay rates to levels that can be probed in future quests by BESIII and LHCb and at the proposed Super Tau-Charm Factory. Both scenarios require significant fine-tuning.
\end{minipage}}

\maketitle

\newpage

\section{Introduction\label{intro}}

The nonleptonic decays of light hyperons that modify the strangeness quantum number by two units have very small rates in the standard model (SM).
Hence such \,$\Delta S$\,=\,2\, processes could serve as an environment in which to search for hints of new physics beyond the SM.
This was first investigated in Ref.\,\cite{He:1997bs}, focusing on the $\mathit\Xi$, which has spin 1/2, turning into a nucleon $N$ and a~pion~$\pi$.
In the spin-3/2 sector, the $\mathit\Omega^-$ hyperon can also be used to test for new physics in \,$\Delta S$\,=\,2 interactions, via the decays \,$\mathit\Omega^-\to n K^-, \mathit\Lambda\pi^-, \mathit\Sigma\pi,\mathit\Sigma^*\pi$.\,

The latest quests for \,$\mathit\Xi\to N\pi$\, were conducted decades ago \cite{Biagi:1982eu,HyperCP:2005jyj} and came up empty, implying the branching-fraction bounds \,${\cal B}(\mathit\Xi^0\to p\pi^-)_{\rm exp}^{}<8\times10^{-6}$\, and \,${\cal B}(\mathit\Xi^-\to n\pi^-)_{\rm exp}^{}<1.9\times10^{-5}$\,~\cite{ParticleDataGroup:2022pth} both at 90\% confidence level (CL).
In the $\mathit\Omega^-$ case, only \,$\mathit\Omega^-\to\mathit\Lambda\pi^-$\, has been searched for \cite{HyperCP:2005jyj}, also with a null outcome, which translated into \,${\cal B}(\mathit\Omega^-\to\mathit\Lambda\pi^-)_{\rm exp}^{}<2.9\times10^{-6}$\, at 90\% CL~\cite{ParticleDataGroup:2022pth}.
As these results are far above the SM expectations, by up to ten orders of magnitude, the window to discover new physics in such \,$\Delta S$\,=\,2\, decays is wide open.
Efforts to pursue this may be made in ongoing experiments, such as LHCb and BESIII.
The former, short of discovery, could better the preceding limits by 3 to 4 orders of magnitude after upcoming upgrades~\cite{AlvesJunior:2018ldo}.
At $e^+e^-$ facilities, BESIII~\cite{BESIII:2020nme} might be able to improve on the $\mathit\Xi$ bounds, and farther in the future  the Super Tau-Charm Factory~\cite{Achasov:2023gey} would expectedly have much enhanced sensitivity to both the $\mathit\Xi$ and $\mathit\Omega^-$ channels~\cite{HBL}.
All this has prompted us to revisit these rare processes in hopes to learn new information about them.

There are relations among several of them, and we identify the independent ones here.
For the hyperons in the octet of ground-state spin-1/2 baryons, the \,$\Delta S$\,=\,2\, nonleptonic decays into two-body final states that are kinematically allowed are \,$\mathit\Xi^0\to p\pi^-,n\pi^0$,\, and \,$\mathit\Xi^-\to n\pi^-$.\,
Within or beyond the SM, the leading operators contributing to these flavor-changing neutral-current processes are of dimension six and consist of four light-quark fields, which can only be the down-type ones.
Thus, the operators entail the conversion of two $s$-quarks into two $d$-quarks, altering isospin by \,$\Delta I$\,=\,1.\,
It follows that, in light of isospin symmetry of the strong interactions, the invariant amplitudes for  $\mathit\Xi\to N\pi$\, satisfy
\begin{align} \label{isospin-x}
\sqrt2\, {\cal M}_{\mathit\Xi^0\to n\pi^0}^{} + {\cal M}_{\mathit\Xi^0\to p\pi^-}^{} + {\cal M}_{\mathit\Xi^-\to n\pi^-}^{} & \,=\, 0 \,. &
\end{align}
As a consequence, it suffices to examine the amplitudes for just two of them, which we choose to be \,$\mathit\Xi^0\to p\pi^-$\, and \,$\mathit\Xi^-\to n\pi^-$.\,

In the decuplet of ground-state spin-3/2 baryons, only the $\mathit\Omega^-$ undergoes predominantly weak decay.
The final states of its $\Delta S$\,=\,2 nonleptonic two-body modes are \,$nK^-,\mathit\Lambda\pi^-,\mathit\Sigma^0\pi^-$, and $\mathit\Sigma^-\pi^0$,\, as well as \,$\mathit\Sigma^{*0}\pi^-$ and $\mathit\Sigma^{*-}\pi^0$,\, the $\mathit\Sigma^*\equiv\mathit\Sigma(1385)$ resonances being also members of the decuplet.
The amplitudes for \,$\mathit\Omega^-\to\mathit\Sigma^0\pi^-,\mathit\Sigma^-\pi^0$\, obey the isospin relation
\begin{align} \label{isospin-o}
{\cal M}_{\mathit\Omega^-\to\mathit\Sigma^0\pi^-}^{} + {\cal M}_{\mathit\Omega^-\to\mathit\Sigma^-\pi^0}^{} & \,=\, 0 \,, &
\end{align}
and so we need not discuss the latter.
The same can be said of \,$\mathit\Omega^-\to\mathit\Sigma^{*0}\pi^-,\mathit\Sigma^{*-}\pi^0$.

The structure of the paper is as follows.
In Sec.\,\ref{SM} we address the \,$\Delta S$\,=\,2\, nonleptonic hyperon decays (NLHD) within the SM.
Specifically, we start by updating the short-distance predictions for \,$\mathit\Xi\to N\pi$\, and subsequently treat their $\mathit\Omega^-$ counterparts.
Moreover, we explicitly look at long-distance effects brought about by \,$\Delta S$\,=\,1\, operators acting twice, which turn out to be numerically important.
Since these processes have relatively low rates already, we do not consider modes with three or more particles in the final states, which have less phase-space.
Beyond the SM, in Secs.\,\,\ref{Z'} and \ref{LQ} we explore how a $Z'$ boson and leptoquarks, respectively, may give rise to substantially amplified contributions to the \,$\Delta S$\,=\,2\, NLHD.
We present our conclusions in Sec.\,\ref{concl}.
In three appendices we summarize the numerical values we use for input parameters, collect the rate formulas for the $\mathit\Omega^-$ modes, and provide further details of the $Z^\prime$ model.

\section{\boldmath$\Delta S$\,=\,2 nonleptonic hyperon decays in the standard model\label{SM}}

\subsection{Short-distance contributions\label{SM-SD}}

In the SM the effective Hamiltonian for \,$\Delta S$\,=\,2\, transitions among light quarks is approximately given by~\cite{Buchalla:1995vs}
\begin{align} \label{smHdsds}
{\cal H}_{\Delta S=2}^{\textsc{sm}} & \,=\, \frac{\eta_{cc}^{} G_{\rm F}^2 m_c^2}{4\pi^2} \big(V_{cd}^*V_{cs}^{}\big){}^2\, {\cal Q}_{LL}^{} \,, &
\end{align}
which involves a QCD-correction factor $\eta_{cc}^{}$, the Fermi constant $G_{\rm F}$, the charm-quark mass $m_c$, the elements $V_{mn}$ of the Cabibbo-Kobayashi-Maskawa (CKM) matrix, and
\begin{align} \label{QLL}
{\cal Q}_{LL}^{} & \,=\, \overline d\gamma^\alpha P_L^{}s\, \overline d\gamma_\alpha^{}P_L^{}s \,=\,
\textit{\textsf t}_{kl,no}^{}\, \overline{\psi_k^{}}\gamma^\alpha P_L^{}\psi_n^{}\, \overline{\psi_l^{}}\gamma_\alpha^{}P_L^{}\psi_o^{} \,, &
\end{align}
with \,$P_L=(1-\gamma_5)/2$,\, the subscripts \,$k,l,n,o=1,2,3$\, being implicitly summed over, \,$\textit{\textsf t}_{kl,no}=0$ except for \,$\textit{\textsf t}_{22,33}=1$, and the light-quark fields \,$\psi_{1,2,3}=u,d,s$.\,
In Eq.\,(\ref{smHdsds}) we have retained only the charm-quark portion, as it dominates the SM short-distance (SD) predictions for the hyperon decays of interest and the neutral-kaon mass difference \,$\Delta M_K^{} = {\rm Re}\langle K^0|{\cal H}_{\Delta S=2}|\bar K^0\rangle/m_{K^0}$,\, the correction from the top and charm-top contributions being merely at the percent level~\cite{Buchalla:1995vs,Brod:2011ty}.

To deal with the hyperon amplitudes generated by ${\cal H}_{\Delta S=2}^{\textsc{sm}}$ requires the hadronized form of ${\cal Q}_{LL}$.
It transforms like $(27_L,1_R)$ under the chiral-symmetry group  ${\rm SU}(3)_L\times{\rm SU}(3)_R$\, and has a leading-order hadronic realization~\cite{He:1997bs,AbdEl-Hady:1998qww} expressible as
\begin{align} \label{OLL}
{\cal O}_{LL}^{} & \,=\, \Lambda_\chi^{} f_\pi^2\, \textit{\textsf t}_{kl,no}^{} \Big[ \hat\beta_{27}^{}\, \big( \xi\overline B\xi^\dagger \big)_{nk\,} \big(\xi B\xi^\dagger\big)_{ol} + \hat\delta_{27}^{}\, \xi_{nx}^{} \xi_{oz}^{} \xi_{vk}^\dagger \xi_{wl}^\dagger\, \big(\overline T_{rvw}\big){}^\alpha (T_{rxz})_\alpha^{}
\Big] \,, &
\end{align}
where $\Lambda_\chi$ is the scale of chiral-symmetry breaking, $f_\pi^{}$ denotes the pion decay constant, $\hat\beta_{27}$ and $\hat\delta_{27}$ are parameters to be fixed below, $B$ and $\xi$ stand for 3$\times$3 matrices incorporating the fields of the lowest-mass octet-baryons and -mesons, respectively, \,$r,v,w,x,z=1,2,3$\, are also summed over, and $(T_{rvw})^\alpha$ is a Rarita-Schwinger field~\cite{Rarita:1941mf} for the spin-3/2 decuplet baryons and has completely symmetric SU(3) indices $(r,v,w)$, the components being explicitly listed in Ref.\,\cite{AbdEl-Hady:1998qww}.
Under \,${\rm SU}(3)_L\times{\rm SU}(3)_R$  rotations \,$B\to\hat UB\hat U^\dagger$,\, $\xi\to\hat L\xi\hat U^\dagger=\hat U\xi\hat R^\dagger$,  and \,$(T_{rvw})^\alpha\to\hat U_{rn}\hat U_{vx}\hat U_{wz}(T_{nxz})^\alpha$,\, where \,$\hat U\in{\rm SU}(3)$\, is implicitly defined by the $\xi$ equation, \,$\hat L\in{\rm SU}(3)_L$,  and \,$\hat R\in{\rm SU}(3)_R$.\,
We will take \,$\Lambda_\chi=4\pi f_\pi^{}$,\, in line with naive-dimensional-analysis arguments~\cite{Manohar:1983md,Georgi:1986kr}.
Note that Eq.\,(\ref{OLL}) does not contain a term directly connecting the decuplet and octet baryons because it is necessarily of higher order in the chiral expansion, needing one derivative of the $\xi$ or $\xi^\dagger$ matrix to contract the Lorentz index in $T^\alpha$~\cite{Tandean:1998ch}, as \,$\gamma_\alpha^{}T^\alpha=\partial_\alpha^{}T^\alpha=0$\, \cite{Rarita:1941mf,Jenkins:1991es}.

\begin{figure}[b]
\includegraphics[trim=43mm 248mm 44mm 37mm,clip,width=0.97\textwidth]{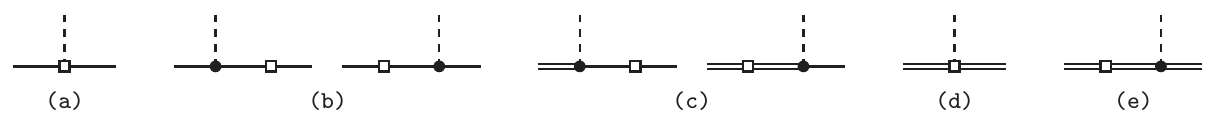} \vspace{-1ex}
\caption{Feynman diagrams for the SM short-distance contributions to (a) \texttt S-wave and (b) \texttt P-wave \,$\mathit\Xi\to N\pi$,\, (c) \texttt P-wave \,$\mathit\Omega^-\mbox{\small$\,\to\mathfrak B$}\phi$,\, and (d) \texttt S-wave and (e) \texttt P-wave \,$\mathit\Omega^-\to\mathit\Sigma^*\pi$.\, Each hollow square symbolizes a~coupling induced by ${\cal H}_{\Delta S=2}^{\textsc{sm}}$ in Eq.\,(\ref{smHdsds}). Here and in Fig.\,\ref{ld-diagrams}, a dashed line represents a pseudoscalar meson, a single (double) solid-line a spin-1/2 (spin-3/2) baryon, and a thick dot a strong vertex from ${\cal L}_{\rm s}$ in Eq.\,(\ref{Ls}).\label{sd-diagrams}}
\end{figure}

The amplitude for a spin-1/2 baryon, {\small$\mathfrak B$}, converting into another one, {\small$\mathfrak B'$}, and a pion can be put in the general form \,$i{\cal M}_{\mathfrak B\to\mathfrak B'\pi} = \bar u_{\mathfrak B'} ({\mathbb A}_{\mathfrak{BB}'} - \gamma_5{\mathbb B}_{\mathfrak{BB}'}) u_{\mathfrak B}$\, comprising, in succession, parity-odd \texttt S-wave and parity-even \texttt P-wave portions~\cite{ParticleDataGroup:2022pth}.
For the former in the \,$\Delta S$\,=\,2\, case, ${\cal H}_{\Delta S=2}^{\textsc{sm}}$ in Eq.\,(\ref{smHdsds}) with ${\cal Q}_{LL}$ changed to ${\cal O}_{LL}$ brings about the diagram depicted in Fig.\,\ref{sd-diagrams}\,(a), leading to~\cite{He:1997bs}
\begin{align} \label{sdA}
{\mathbb A}_{\mathit\Xi^0 p}^{\scriptscriptstyle\rm(SM,SD)} & \,=\, {\mathbb A}_{\mathit\Xi^-n}^{\scriptscriptstyle\rm(SM,SD)} \,=\, \frac{{\textsc c}_{\textsc{sm}}}{\sqrt2} \,, &
\end{align}
where \,${\textsc c}_{\textsc{sm}}=\eta_{cc}^{} G_{\rm F}^2 m_c^2\, \big(V_{cd}^*V_{cs}^{}\big){}^2 f_\pi^2 \hat\beta_{27}/\pi$.\,
The corresponding $\mathbb B$ pieces are calculated from pole diagrams, displayed in Fig.\,\ref{sd-diagrams}\,(b), which depend on ${\textsc c}_{\textsc{sm}}$ and also have a vertex furnished by the leading-order strong-interaction chiral Lagrangian~\cite{Jenkins:1991es,Bijnens:1985kj}
\begin{align} \label{Ls}
{\cal L}_{\rm s} & \,\supset\, {\rm Tr}\big(D\,\overline B\gamma^\alpha\gamma_5^{}\{{\cal A}_\alpha,B\} + F\,\overline B\gamma^\alpha\gamma_5^{}[{\cal A}_\alpha,B]\big) \,+\, {\mathscr H}\, \big(\raisebox{-1pt}{$\overline T_{klv}$}\big)^\alpha\, \gamma^\mu\gamma_5^{}\, ({\cal A}_{vw})_\mu^{}\, (T_{klw})_\alpha^{}
\nonumber \\ & ~~~~ +\, \epsilon_{kln}^{}\, {\cal C}\, \Big[ \big(\raisebox{-1pt}{$\overline B$}\big)_{kv}\, ({\cal A}_{lw})_\alpha^{}\, (T_{nvw})^\alpha + \big(\raisebox{-1pt}{$\overline T_{nvw}$}\big)^\alpha\, ({\cal A}_{wl})_\alpha^{}\, B_{vk} \Big] \,, &
\end{align}
where $D$, $F$, $\mathscr H$, and $\cal C$ are constants and \,${\cal A}_\alpha^{} = i\big(\xi\partial_\alpha^{}\xi^\dagger - \xi^\dagger\partial_\alpha^{}\xi\big)/2$.\,
The results are~\cite{He:1997bs}
\begin{align} \label{sdB}
{\mathbb B}_{\mathit\Xi^0p}^{\scriptscriptstyle\rm(SM,SD)} & \,=\, \frac{D+F}{\sqrt2} \bigg(\frac{m_N+m_{\mathit\Xi}}{m_{\mathit\Xi}-m_N}\bigg) {\textsc c}_{\textsc{sm}} \,, &
{\mathbb B}_{\mathit\Xi^-n}^{\scriptscriptstyle\rm(SM,SD)} & \,=\, \frac{D-F}{\sqrt2} \bigg(\frac{m_N+m_{\mathit\Xi}}{m_N-m_{\mathit\Xi}}\bigg) {\textsc c}_{\textsc{sm}} \,, &
\end{align}
where $m_N^{}$ and $m_{\mathit\Xi}^{}$ are isospin-averaged nucleon and $\mathit\Xi^{-,0}$ masses, respectively.

The mode \,$\mathit\Omega^-${\small$\to\mathfrak B$}$\phi$,\, with $\phi$ being a pseudoscalar meson, is made up of \texttt P-wave and \texttt D-wave transitions.
In the SM, the SD contribution to the former proceeds from the pole diagrams exhibited in Fig.\,\ref{sd-diagrams}\,(c) which include not only a weak coupling produced by ${\cal H}_{\Delta S=2}^{\textsc{sm}}$ but also a strong vertex from the $\cal C$ term of Eq.\,(\ref{Ls}).
The \texttt D-wave piece arises from a higher order in the chiral expansion and hence will be neglected.
Writing the amplitude accordingly as  \,$i{\cal M}_{\mathit\Omega^-\to\mathfrak B\phi}={\mathbb C}_{\mathfrak B\phi}\,\bar u_{\mathfrak B}\,u_{\mathit\Omega}^\alpha\, \tilde p_\alpha^{}$,\,  with $\tilde p$ being the four-momentum of $\phi$, we then have
\begin{align} \label{sdO}
{\mathbb C}_{nK^-}^{\scriptscriptstyle\rm(SM,SD)} & \,=\, \frac{\cal C}{\sqrt2} \bigg[ \frac{{\textsc c}_{\textsc{sm}}}{m_{\mathit\Xi}-m_N} - \frac{\widetilde{\textsc c}{}_{\textsc{sm}}}{3(m_\mathit\Omega-m_{\mathit\Sigma^*})} \bigg] \,, &
\nonumber \\
{\mathbb C}_{\mathit\Lambda\pi^-}^{\scriptscriptstyle\rm(SM,SD)} & \,=\, \frac{{\cal C}\, \widetilde{\textsc c}{}_{\textsc{sm}}}{2\sqrt3\, (m_\mathit\Omega-m_{\mathit\Sigma^*})} \,, \hspace{7em}
{\mathbb C}_{\mathit\Sigma^0\pi^-}^{\scriptscriptstyle\rm(SM,SD)} \,=\, \frac{-{\cal C}\,\widetilde{\textsc c}{}_{\textsc{sm}}}{6(m_\mathit\Omega-m_{\mathit\Sigma^*})} \,,
\end{align}
where \,$\widetilde{\textsc c}{}_{\textsc{sm}}=\eta_{cc}^{} G_{\rm F}^2 m_c^2\, \big(V_{cd}^*V_{cs}^{}\big){}^2 f_\pi^2 \hat\delta_{27}^{}/\pi$\, and $m_{\mathit\Sigma^*}$ is the isospin-averaged mass of the $\mathit\Sigma(1385)$ resonances.

As for \,$\mathit\Omega^-\to\mathit\Sigma^{*0}\pi^-$,\, it is described by \texttt S-, \texttt P-, \texttt D-, and \texttt F-wave amplitudes.
The first two of them can be expressed as \,$i{\cal M}_{\mathit\Omega^-\to\mathit\Sigma^{*0}\pi^-}=\bar u_{\mathit\Sigma^*}^\alpha \big(\tilde{\mbox{\small$\mathbb A$}}_{\mathit\Sigma^*\pi}-\gamma_5^{}\, \tilde{\mbox{\small$\mathbb B$}}_{\mathit\Sigma^*\pi}\big) u_{\mathit\Omega,\alpha}$,\, and in the SM the SD ones are determined from the leading-order diagrams in Fig.\,\ref{sd-diagrams}\,(d,e), respectively.
Thus, we find
\begin{align} \label{sdOSs}
\tilde{\mbox{\small$\mathbb A$}}_{\mathit\Sigma^*\pi}^{\scriptscriptstyle\rm(SM,SD)} & \,=\, \frac{\widetilde{\textsc c}{}_{\textsc{sm}}}{\sqrt3} \,, &
\tilde{\mbox{\small$\mathbb B$}}_{\mathit\Sigma^*\pi}^{\scriptscriptstyle\rm(SM,SD)} & \,=\, \frac{-\mathscr H}{3\sqrt3} \bigg(\frac{m_{\mathit\Omega}+m_{\mathit\Sigma^*}}{m_{\mathit\Omega}-m_{\mathit\Sigma^*}}\bigg) \widetilde{\textsc c}{}_{\textsc{sm}} \,. &
\end{align}
The \texttt D- and \texttt F-wave terms occur at higher chiral orders and will therefore be ignored.

The value of $\hat\beta_{27}$ can be inferred, with the aid of flavor-SU(3) symmetry, from the \,$\Delta I$\,=\,3/2  amplitudes for the measured $\Delta S$\,=\,1 NLHD.
This is in analogy to linking the matrix elements for $K^0$-$\bar K^0$ mixing and the \,$\Delta I$\,=\,3/2\, component of $K\to2\pi$\, decay~\cite{Donoghue:1982cq}.
In the SM the pertinent  $\Delta S$\,=\,1\, Hamiltonian at short distance is
\begin{align} \label{DS=1,DI=3/2}
{\cal H}_{\Delta I=3/2,\Delta S=1}^{\textsc{sm}} & \,=\, \sqrt8\, (\hat{\textsc c}_1+\hat{\textsc c}_2)\, G_{\rm F}^{}\, V_{ud}^*V_{us}^{}\,
{\cal Q}_{\scriptscriptstyle\Delta S=1}^{\scriptscriptstyle\Delta I=3/2} \,, &
\end{align}
where $\hat{\textsc c}_{1,2}$ designate the main Wilson coefficients and
\,${\cal Q}_{\scriptscriptstyle\Delta S=1}^{\scriptscriptstyle\Delta I=3/2} = \tilde t_{kl,no}^{}\, \overline{\psi_k^{}}\gamma^\alpha P_L^{}\psi_n^{}\, \overline{\psi_l^{}}\gamma_\alpha^{}P_L^{}\psi_o^{}$,\,
with  $\tilde t_{kl,no}=0$\, except for \,$\tilde t_{12,13}=\tilde t_{12,31}=\tilde t_{21,13}=\tilde t_{21,31}=-\tilde t_{22,23}=-\tilde t_{22,32}=1/6$.\,
This operator also transforms as $(27_L,1_R)$ under \,${\rm SU}(3)_L\times{\rm SU}(3)_R$.\,
Accordingly, the hadronic realization of ${\cal Q}_{\scriptscriptstyle\Delta S=1}^{\scriptscriptstyle\Delta I=3/2}$ at lowest order in the chiral expansion is~\cite{He:1997bs,AbdEl-Hady:1998qww}
\begin{align} \label{ODI=3/2}
{\cal O}_{\scriptscriptstyle\Delta S=1}^{\scriptscriptstyle\Delta I=3/2} & \,=\, \Lambda_\chi^{} f_\pi^2\, \tilde t_{kl,no}^{} \Big[ \hat\beta_{27}^{}\, \big( \xi\overline B\xi^\dagger \big)_{nk}\, \big(\xi B\xi^\dagger\big)_{ol} + \hat\delta_{27}^{}\, \xi_{nx}^{} \xi_{oz}^{} \xi_{vk}^\dagger \xi_{wl}^\dagger\, \big(\overline T_{rvw}\big){}^\eta\, (T_{rxz})_\eta^{}
\Big] \,. &
\end{align}
Since experiments reveal that the \,$\Delta S$\,=\,1\, NLHD are dominated by their $(8_L,1_R)$ amplitudes, which are {\small\,$\sim$\,}20 times bigger in size than their $(27_L,1_R)$ counterparts, theoretical examination of the latter suffers from large uncertainties because of complications due to isospin-mixing effects plus the ambiguity associated with the \texttt S-wave/\texttt P-wave problem for the spin-1/2 hyperons~\cite{Maltman:1995qw,Na:1997am}.
Nevertheless, there is one exception, namely that the \texttt S-wave amplitude for \,$\mathit\Sigma^+\to n\pi^+$  receives no $(8_L,1_R)$ contribution in chiral perturbation theory up to second order in external momentum or meson mass~\cite{Bijnens:1985kj,Borasoy:1998ku} and therefore offers possibly the cleanest way to assess $\hat\beta_{27}$.
From \,$\mathit\Sigma^+\to n\pi^+$  measurements~\cite{ParticleDataGroup:2022pth}, we get \,${\mathbb A}_{\mathit\Sigma^+n}^{\scriptscriptstyle\rm(exp)}=1.40(27)\times10^{-8}$.\,
From Eq.\,(\ref{DS=1,DI=3/2}) with ${\cal Q}_{\scriptscriptstyle\Delta S=1}^{\scriptscriptstyle\Delta I=3/2}$ replaced by ${\cal O}_{\scriptscriptstyle\Delta S=1}^{\scriptscriptstyle\Delta I=3/2}$, we derive \,${\mathbb A}_{\mathit\Sigma^+n}^{\scriptscriptstyle\rm(theory)}=(\hat{\textsc c}_1+\hat{\textsc c}_2) G_{\rm F}^{} V_{ud}^*V_{us}^{} \Lambda_\chi f_\pi^{} \hat\beta_{27}$.
Equating these ${\mathbb A}$s, assuming that higher chiral orders can be neglected, and using \,$0.64\le\hat{\textsc c}_1+\hat{\textsc c}_2\le0.72$\, computed in Ref.\,\cite{Buchalla:1995vs} at leading order (for the renormalization scale of 1 GeV and QCD scales of 215-435 MeV) and the $f_\pi^{}$, $G_{\rm F}$, and $V_{ud}^*V_{us}^{}$ values collected in Appendix \ref{num}, we then extract
\begin{align} \label{hh27}
\hat\beta{}_{27}^{} & \,=\, 0.076(15) \,. &
\end{align}

As for $\hat\delta_{27}$, at the moment it cannot be estimated unambiguously from experiment because its role in the observed $\mathit\Omega^-$ transitions is minor compared to those of the $(8_L,1_R)$ parameters.
Since, like $\hat\beta_{27}$, it belongs to $(27_L,1_R)$ interactions, to illustrate how $\hat\delta_{27}$ may influence the $\mathit\Omega^-$ channels of interest, we will set \,$\hat\delta_{27}=\hat\beta_{27}$ \,or\, $-\hat\beta_{27}$.\,

From ${\cal M}_{\mathfrak B\to\mathfrak B'\pi}$ follows the rate \,$\Gamma_{\mathfrak B\to\mathfrak B'\pi}=|\mbox{\small\bf p}'|\big[|{\mathbb A}_{\mathfrak{BB}'}|^2(\texttt E'+m_{\mathfrak B'})+|{\mathbb B}_{\mathfrak{BB}'}|^2(\texttt E'-m_{\mathfrak B'})\big]/(4\pi m_{\mathfrak B})$,  where {\small\bf p}$'$ ($\texttt E'$) is the three-momentum (energy) of {\small$\mathfrak B'$} in the rest frame of {\small$\mathfrak B$}.
We can employ this to evaluate the contributions of Eqs.\,\,(\ref{sdA}) and (\ref{sdB}) to \,$\mathit\Xi\to N\pi$,\, with the central values of $\hat\beta_{27}$ above and of $\eta_{cc}^{}$, $m_c^{}$, $|V_{cd}V_{cs}|$, $D$, and $F$ quoted in Appendix\,\,\ref{num}, leading to the branching fractions
\begin{align} \label{smsdBXi2Npi}
{\cal B}\big(\mathit\Xi^0\to p\pi^-\big){}_{\textsc{sm}}^{\textsc{sd}} & \,=\, 3.0\times10^{-16} \,, &
{\cal B}\big(\mathit\Xi^0\to n\pi^0\big){}_{\textsc{sm}}^{\textsc{sd}} & \,=\, 3.0\times10^{-16} \,, & \nonumber \\
{\cal B}\big(\mathit\Xi^-\to n\pi^-\big){}_{\textsc{sm}}^{\textsc{sd}} & \,=\, 7.9\times10^{-17} \,.
\end{align}

For the $\mathit\Omega^-$ transitions, from the aforementioned amplitudes it is straightforward to obtain the rates written in Eqs.\,(\ref{GO2Bphi})-(\ref{GO2Sspi}).
With Eqs.\,(\ref{sdO})-(\ref{sdOSs}) and the central values of the input parameters, including $\cal C$ and $\mathscr H$ from Appendix\,\,\ref{num}, we then find
\begin{align} \label{smsdBO2Bphi}
{\cal B}\big(\mathit\Omega^-\to nK^-\big){}_{\textsc{sm}}^{\textsc{sd}}         & \,=\, (1.4,9.4)\times10^{-17} \,, &
{\cal B}\big(\mathit\Omega^-\to\mathit\Lambda\pi^-\big){}_{\textsc{sm}}^{\textsc{sd}}  & \,=\, 2.0\times10^{-17} \,, &
\nonumber \\
{\cal B}\big(\mathit\Omega^-\to\mathit\Sigma^0\pi^-\big){}_{\textsc{sm}}^{\textsc{sd}} & \,=\, 4.6\times10^{-18} \,, &
{\cal B}\big(\mathit\Omega^-\to\mathit\Sigma^{*0}\pi^-\big){}_{\textsc{sm}}^{\textsc{sd}} & \,=\, 2.8\times10^{-17} \,,
\end{align}
where the two entries for \,$\mathit\Omega^-\to nK^-$\, correspond to \,$\hat\delta_{27}^{}=(1,-1)\hat\beta_{27}^{}$,\, respectively.

\subsection{Long-distance contributions\label{SM-LD}}

\begin{figure}[t]
\includegraphics[trim=45mm 225mm 44mm 38mm,clip,width=0.92\textwidth]{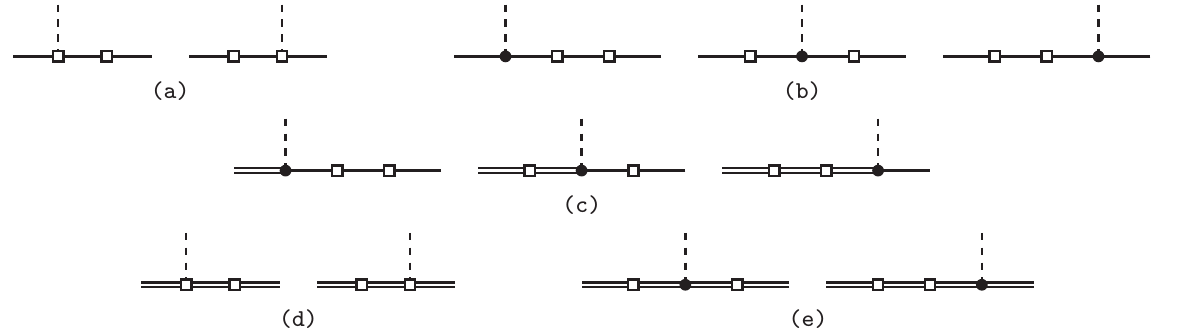} \vspace{-7pt}
\caption{Feynman diagrams for the SM long-distance contributions to (a) \texttt S-wave and (b) \texttt P-wave \,$\mathit\Xi\to N\pi$,\, (c) \texttt P-wave \,$\mathit\Omega^-\mbox{\small$\,\to\mathfrak B$}\phi$,\, and (d) \texttt S-wave and (e) \texttt P-wave \,$\mathit\Omega^-\to\mathit\Sigma^*\pi$.\,
Each hollow square symbolizes a weak coupling supplied by ${\cal L}_{\scriptscriptstyle\Delta S=1}^{\textsc{sm}}$ in Eq.\,(\ref{Lw}).\label{ld-diagrams}}
\end{figure}

These $\mathit\Xi$ and $\mathit\Omega^-$ modes are also affected by the pole diagrams depicted in Fig.\,\ref{ld-diagrams}, with two couplings from the lowest-order \,$\Delta S$\,=\,1\, chiral Lagrangian~\cite{Bijnens:1985kj,Jenkins:1991bt}
\begin{align} \label{Lw}
{\cal L}_{\scriptscriptstyle\Delta S=1}^{\textsc{sm}} & \,=\, {\rm Tr}\big( h_D^{}\, \overline B\big\{ \xi^\dagger\hat\kappa\xi,B \big\} + h_F^{}\, \overline B\big[\xi^\dagger\hat\kappa\xi,B\big] \big) + h_C^{}\, \big(\overline T_{kln}\big)^\eta\, \big(\xi^\dagger\hat\kappa\xi\big)_{no}\, (T_{klo})_\eta^{} \,, &
\end{align}
which transforms as $(8_L,1_R)$ under \,${\rm SU}(3)_L\times{\rm SU}(3)_R$\, and contains parameters $h_{D,F,C}$ and a 3$\times$3 matrix $\hat\kappa$ with elements \,$\hat\kappa_{kl}=\delta_{2k}\delta_{3l}$.
The diagrams for the $\mathbb B$s and $\mathbb C$s, in Fig.\,\ref{ld-diagrams}\,(b,c,e), again include a strong vertex from Eq.\,(\ref{Ls}) as well.
Accordingly, for \,$\mathit\Xi\to N\pi$\, we derive the long-distance (LD) contributions
\begin{align} \label{ldA}
{\mathbb A}_{\mathit\Xi^0p}^{\scriptscriptstyle\rm(SM,LD)} & \,=\, \frac{1}{\sqrt2\, f_\pi^{}} \Bigg[ \frac{h_D^2-h_F^2}{m_N-m_{\mathit\Sigma}} + \frac{h_D^2-h_F^2}{2(m_{\mathit\Sigma}-m_{\mathit\Xi})} + \frac{h_D^2-9h_F^2}{6(m_{\mathit\Xi}-m_{\mathit\Lambda})} \Bigg] \,, &
\nonumber \\
{\mathbb A}_{\mathit\Xi^-n}^{\scriptscriptstyle\rm(SM,LD)} & \,=\, \frac{1}{\sqrt2\, f_\pi^{}} \Bigg[ \frac{h_D^2-h_F^2}{m_{\mathit\Xi}-m_{\mathit\Sigma}} + \frac{h_D^2-h_F^2}{2(m_{\mathit\Sigma}-m_N)} + \frac{h_D^2-9h_F^2}{6(m_N-m_{\mathit\Lambda})} \Bigg] \,,
\end{align}
\begin{align} \label{ldB}
{\mathbb B}_{\mathit\Xi^0p}^{\scriptscriptstyle\rm(SM,LD)} & \,=\,
\frac{h_D-h_F}{\sqrt2\, f_\pi} \bigg(\frac{m_{\mathit\Xi}+m_N}{m_{\mathit\Sigma}-m_N}\bigg) \bigg[
\frac{D(h_D-3h_F)}{3(m_{\mathit\Xi}-m_{\mathit\Lambda})} - F\, \frac{h_D+h_F}{m_{\mathit\Xi}-m_{\mathit\Sigma}} \bigg]
\nonumber \\ & ~~~ +\, \frac{D+F}{2\sqrt2\, f_\pi} \bigg(\frac{m_{\mathit\Xi}+m_N}{m_{\mathit\Xi}-m_N}\bigg) \Bigg[
\frac{h_D^2-9h_F^2}{3(m_{\mathit\Xi}-m_{\mathit\Lambda})} + \frac{h_D^2-h_F^2}{m_{\mathit\Xi}-m_{\mathit\Sigma}} \Bigg] \,, &
\nonumber \\ \raisebox{5ex}{}
{\mathbb B}_{\mathit\Xi^-n}^{\scriptscriptstyle\rm(SM,LD)} & \,=\,
\frac{h_D+h_F}{\sqrt2\, f_\pi} \bigg(\frac{m_N+m_{\mathit\Xi}}{m_{\mathit\Sigma}-m_{\mathit\Xi}}\bigg) \bigg[
\frac{D(h_D+3h_F)}{3(m_N-m_{\mathit\Lambda})} + F\, \frac{h_D-h_F}{m_N-m_{\mathit\Sigma}} \bigg]
\nonumber \\ & ~~~ +\, \frac{D-F}{2\sqrt2\, f_\pi} \bigg(\frac{m_N+m_{\mathit\Xi}}{m_N-m_{\mathit\Xi}}\bigg) \Bigg[
\frac{h_D^2-9h_F^2}{3(m_N-m_{\mathit\Lambda})} + \frac{h_D^2-h_F^2}{m_N-m_{\mathit\Sigma}} \Bigg] \,,
\end{align}
where $m_{\mathit\Sigma}^{}$ is the average of the $\mathit\Sigma^{+,0,-}$ masses, and for the $\mathit\Omega^-$ channels
\begin{align} \label{ldO}
{\mathbb C}_{nK^-}^{\scriptscriptstyle\rm(SM,LD)} & \,=\, \frac{{\cal C}\, h_C}{6\sqrt2\, f_\pi^{}\, (m_\mathit\Omega-m_{\mathit\Xi^*})} \bigg[\frac{h_D+3 h_F}{m_{\mathit\Lambda}-m_N} - \frac{h_D-h_F}{m_{\mathit\Sigma}-m_N} + \frac{4 h_C}{3(m_\mathit\Omega-m_{\mathit\Sigma^*})}\bigg]
\nonumber \\ & ~~~~ -\, \frac{\cal C}{2\sqrt2\, f_\pi^{}\, (m_{\mathit\Xi}-m_N)} \Bigg[ \frac{h_D^2-9 h_F^2}{3(m_{\mathit\Lambda}-m_N)} + \frac{h_D^2-h_F^2}{m_{\mathit\Sigma}-m_N} \Bigg] \,,
\nonumber \\ \raisebox{5ex}{}
{\mathbb C}_{\mathit\Lambda\pi^-}^{\scriptscriptstyle\rm(SM,LD)} & \,=\, \frac{{\cal C}\, h_C}{6\sqrt3\, f_\pi^{}\, (m_\mathit\Omega-m_{\mathit\Xi^*})} \bigg(\frac{h_D-3 h_F}{m_{\mathit\Xi}-m_{\mathit\Lambda}} - \frac{2 h_C}{m_\mathit\Omega-m_{\mathit\Sigma^*}}\bigg) \,,
\nonumber \\ \raisebox{5ex}{}
{\mathbb C}_{\mathit\Sigma^0\pi^-}^{\scriptscriptstyle\rm(SM,LD)} & \,=\, \frac{{\cal C}\,h_C}{6 f_\pi^{}\, (m_\mathit\Omega-m_{\mathit\Xi^*})} \bigg[\frac{h_D+h_F}{m_{\mathit\Xi}-m_{\mathit\Sigma}} + \frac{2 h_C}{3(m_\mathit\Omega-m_{\mathit\Sigma^*})}\bigg] \,, &
\\ \raisebox{5ex}{}
\tilde{\mbox{\small$\mathbb A$}}_{\mathit\Sigma^*\pi}^{\scriptscriptstyle\rm(SM,LD)} & \,=\, \frac{h_C^2}{3\sqrt3\, f_\pi^{}} \bigg( \frac{1}{m_{\mathit\Xi^*}-m_{\mathit\Sigma^*}} - \frac{1}{m_{\mathit\Omega}-m_{\mathit\Xi^*}} \bigg) \,, &
\nonumber \\ \raisebox{4ex}{}
\tilde{\mbox{\small$\mathbb B$}}_{\mathit\Sigma^*\pi}^{\scriptscriptstyle\rm(SM,LD)} & \,=\, \frac{\mathscr H\, h_C^2}{9\sqrt3\, f_\pi^{}} \bigg(\frac{m_{\mathit\Omega}+m_{\mathit\Sigma^*}}{m_{\mathit\Omega}-m_{\mathit\Xi^*}}\bigg)\bigg( \frac{1}{m_{\mathit\Xi^*}-m_{\mathit\Sigma^*}} - \frac{2}{m_{\mathit\Omega}-m_{\mathit\Sigma^*}} \bigg) \,,
\end{align}
where $m_{\mathit\Xi^*}$ is the isospin-averaged mass of the $\mathit\Xi(1530)$ resonances, which are of spin-3/2 and also members of the baryon decuplet.

The unknowns here are $h_{D,F,C}$, but they can be evaluated from the available data on the \,$\Delta S$\,=\,1\, processes \,$\mathit\Lambda\to N\pi$,\, $\mathit\Sigma\to N\pi$,\, $\mathit\Xi\to\mathit\Lambda\pi$,\, and \,$\mathit\Omega^-\to\mathit\Lambda K^-,\mathit\Xi\pi$\, \cite{Bijnens:1985kj,Jenkins:1991bt}.
Thus, performing a~least-squares fit of the octet-hyperon \texttt S-wave and $\mathit\Omega^-$ \texttt P-wave decay amplitudes at leading order to their empirical values~\cite{ParticleDataGroup:2022pth} yields the numbers in Eq.\,(\ref{sfit}).
Subsequently, combining the central values of $h_{D,F}$ with Eqs.\,(\ref{ldA})-(\ref{ldB}), we arrive at \,${\cal B}\big(\mathit\Xi^0\to p\pi^-\big){}_{\textsc{sm}}^{\textsc{ld}} = 2.7\times10^{-15}$,\, ${\cal B}\big(\mathit\Xi^0\to n\pi^0\big){}_{\textsc{sm}}^{\textsc{ld}} = 4.8\times10^{-16}$,  and \,${\cal B}\big(\mathit\Xi^-\to n\pi^-\big){}_{\textsc{sm}}^{\textsc{ld}} = 1.5\times10^{-15}$,\, which exceed their SD counterparts in Eq.\,(\ref{smsdBXi2Npi}) by up to \,{\footnotesize$\sim$\,}20 times, implying that we need to put together the LD and SD amplitudes.
Since the relative phase between the two is undetermined, we simply subtract one from the other or add them up to find
\begin{align} \label{smBX2Npi}
{\cal B}\big(\mathit\Xi^0\to p\pi^-\big)_{\textsc{sm}} & \,=\, (2.8,3.1)\times10^{-15} \,, &
{\cal B}\big(\mathit\Xi^0\to n\pi^0\big)_{\textsc{sm}} & \,=\, (1.5,0.02)\times10^{-15} \,, &
\nonumber \\
{\cal B}\big(\mathit\Xi^-\to n\pi^-\big)_{\textsc{sm}} & \,=\, (1.2,1.8)\times10^{-15} \,.
\end{align}
In the case of \,$\mathit\Omega^-\mbox{\small$\,\to\mathfrak B$}\phi$,\, the LD contributions turn out to be significantly bigger than the SD ones, but the two are not highly disparate in \,$\mathit\Omega^-\to\mathit\Sigma^*\pi$,\, similarly to \,$\mathit\Xi\to N\pi$.\,
Explicitly, neglecting the SD ones in \,$\mathit\Omega^-\mbox{\small$\,\to\mathfrak B$}\phi$,\, with the central values of $h_{D,F,C}$ in Eq.\,(\ref{sfit}) we have
\begin{align} \label{smBO2Bphi}
{\cal B}\big(\mathit\Omega^-\to nK^-\big)_{\textsc{sm}}         & \,=\, 3.4\times10^{-13} \,, &
{\cal B}\big(\mathit\Omega^-\to\mathit\Lambda\pi^-\big)_{\textsc{sm}}  & \,=\, 8.2\times10^{-14} \,, & \nonumber \\
{\cal B}\big(\mathit\Omega^-\to\mathit\Sigma^0\pi^-\big)_{\textsc{sm}} & \,=\, 1.5\times10^{-14} \,, &
{\cal B}\big(\mathit\Omega^-\to\mathit\Sigma^{*0}\pi^-\big)_{\textsc{sm}} & \,=\, (2.0,5.4)\times10^{-17} \,,
\end{align}
where the first three results surpass the ones in Eq.\,(\ref{smsdBO2Bphi}) by over 3 orders of magnitude.

Although the preceding $h_{D,F}$ numbers give rise to a good fit to the \texttt S-wave \,$\Delta S$\,=\,1\, NLHD, they translate into a poor representation of the \texttt P waves.
On the other hand, it is possible to come up with a satisfactory account of the \texttt P waves, but end up with a disappointing description of the \texttt S waves.
This is a well-known longstanding problem~\cite{Bijnens:1985kj,Na:1997am,Borasoy:1998ku,Jenkins:1991bt}, which lies beyond the scope of our analysis.
Here we would merely like to see how different possible picks of $h_{D,F,C}$ might alter the  $\Delta S$\,=\,2\, predictions.
Particularly, fitting to the \,$\Delta S$\,=\,1\, octet-hyperon and $\mathit\Omega^-$ \texttt P-waves produces the entries in Eq.\,(\ref{pfit}).
These cause the LD components in \,$\mathit\Xi\to N\pi$\, to be much greater than the SD ones, which now impact the branching fractions by no more than 15\%,
\begin{align} \label{smBX2Npi-v2}
{\cal B}\big(\mathit\Xi^0\to p\pi^-\big)_{\textsc{sm}} & \,=\, (2.85,2.91)\times10^{-13} \,, &
{\cal B}\big(\mathit\Xi^0\to n\pi^0\big)_{\textsc{sm}} & \,=\, (1.0,1.4)\times10^{-14} \,, &
\nonumber \\
{\cal B}\big(\mathit\Xi^-\to n\pi^-\big)_{\textsc{sm}} & \,=\, (1.1,1.2)\times10^{-13} \,,
\end{align}
whereas the $\mathit\Omega^-$ outcomes,
\begin{align} \label{smBO2Bphi-v2}
{\cal B}\big(\mathit\Omega^-\to nK^-\big){}_{\textsc{sm}}^{} & \,=\, 7.5\times10^{-13} \,, & {\cal B}\big(\mathit\Omega^-\to\mathit\Lambda\pi^-\big){}_{\textsc{sm}}^{} & \,=\, 1.3\times10^{-13} \,, &
\nonumber \\
{\cal B}\big(\mathit\Omega^-\to\mathit\Sigma^0\pi^-\big){}_{\textsc{sm}}^{} & \,=\, 5.7\times10^{-15} \,, &
{\cal B}\big(\mathit\Omega^-\to\mathit\Sigma^{*0}\pi^-\big)_{\textsc{sm}} & \,=\, (2.0,4.8)\times10^{-17} \,,
\end{align}
are roughly comparable to those in Eq.\,(\ref{smBO2Bphi}).

\begin{table}[t]
\begin{tabular}{|c||c|c|c|} \hline
\multirow{2}{*}{Mode} & \multicolumn{3}{c|}{Branching fractions} \\ \cline{2-4}
& \footnotesize SD & \footnotesize SD\,+\,LD ($\tilde{\textsc s}$) & \footnotesize SD\,+\,LD ($\tilde{\textsc p}$) \\ \hline
$\mathit\Xi^0\to p\pi^-\vphantom{|_|^{|^|}}$ & $(0.03,1)\times 10^{-15}$ & $(0.01,2.6)\times 10^{-14}$ & $~(0.7,8.2)\times 10^{-13}~$ \\
$\mathit\Xi^0\to n\pi^0\vphantom{|_|^|}$ & $(0.03,1)\times 10^{-15}$ & $(0.,0.9)\times 10^{-15}$ & $~(0.03,0.4)\times 10^{-13}~$ \\
$\mathit\Xi^-\to n\pi^-\vphantom{|_|^|}$ & $(0.07,2.6)\times 10^{-16}$ & ~$(0.01,1.3)\times 10^{-14}$~ & $(0.03,0.3)\times 10^{-12}$\\
$\mathit\Omega^-\to nK^-\vphantom{|_|^|}$ & $(0.1,6.5)\times 10^{-17}$ & $(0.2,0.6)\times 10^{-12}$ & $(0.2,2.1)\times 10^{-12}$ \\
$\mathit\Omega^-\to\mathit\Lambda\pi^-\vphantom{|_|^|}$  & $(0.2,7.1)\times 10^{-17}$   & $(0.4,1.5)\times 10^{-13}$ & $(0.2,4.2)\times 10^{-13}$ \\
$\mathit\Omega^-\to\mathit\Sigma^0\pi^-\vphantom{|_|^|}$ & $~(0.04,1.7)\times 10^{-17}~$ & $~(0.5,3.1)\times 10^{-14}~$ & $(0.05,2.2)\times 10^{-14}$ \\
$~\mathit\Omega^-\to\mathit\Sigma^{*0}\pi^-~\vphantom{|_|^|}$ &$~(0.3,9)\times10^{-17}$ & $~(0.6,7.5)\times10^{-17}$ &$~(1,14)\times10^{-17}$ \\  \hline
\end{tabular}
\caption{The 90\%-CL intervals of branching fractions of \,$\Delta S$\,=\,2\, nonleptonic hyperon decays from the short-distance and complete contributions of the SM, as explained in the text.\label{t:smrates}}
\end{table}

To understand the parametric uncertainty of these SM predictions and their correlations, we quote the 90\%-CL intervals for each observable at a time in Table\,\,\ref{t:smrates}, after implementing the steps outlined in Appendix\,\,\ref{num}.
The second column of the table lists only the SD contributions, with $\hat\delta_{27}$ selected to have the same sign as $\hat\beta_{27}$.
For the third column (labeled $\tilde{\textsc s}$), we have incorporated the LD components, taking them to have the same phase as the SD ones and including the correlations between the values of $h_{D,F,C}$ as obtained from fitting the \texttt S waves of octet-hyperon nonleptonic decays and \texttt P waves of \,$\mathit\Omega^-\mbox{\small$\,\to\mathfrak B$}\phi$\, in the \,$\Delta S$\,=\,1\, sector.
For the fourth column (labeled $\tilde{\textsc p}$), we have repeated this exercise but with $\hat\delta_{27}$ and $\hat\beta_{27}$ having different signs, the SD and LD parts being opposite in phase, and $h_{D,F,C}$ from fitting the \texttt P waves of both the \,$\Delta S$\,=\,1\, octet-hyperon and $\mathit\Omega^-$ decays.
As anticipated, for the last column the SD terms are, on the whole, numerically insignificant relative to the LD ones.

\begin{figure}[t]
\includegraphics[width=65mm]{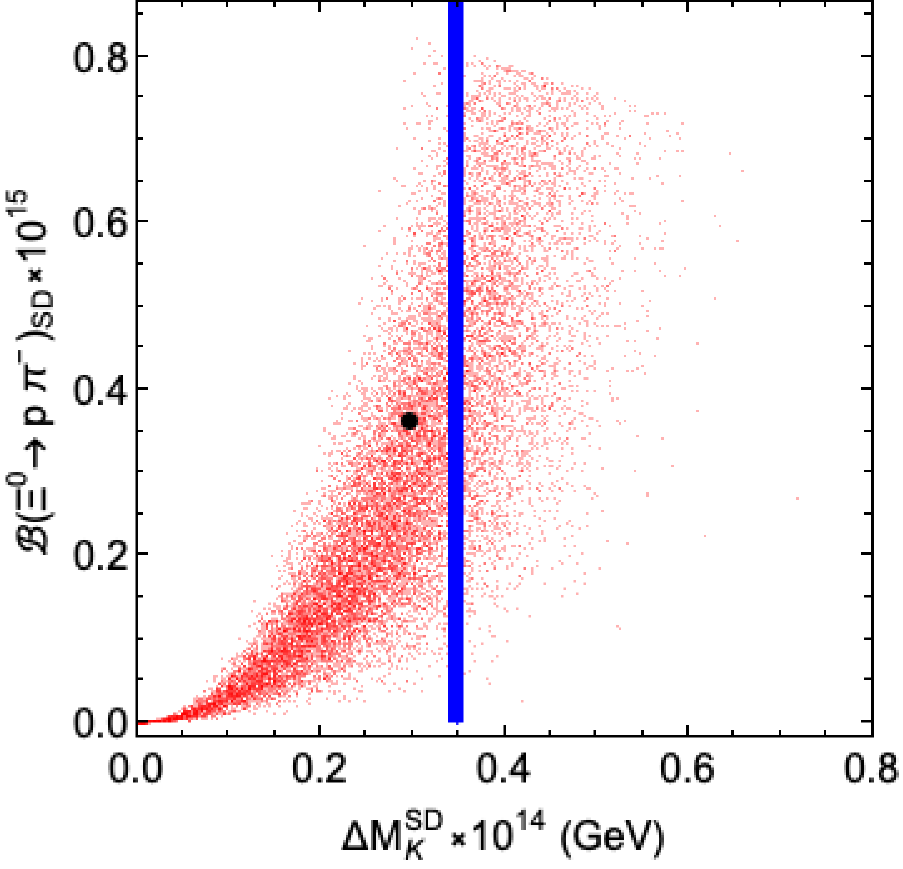} ~ ~ \includegraphics[width=65mm]{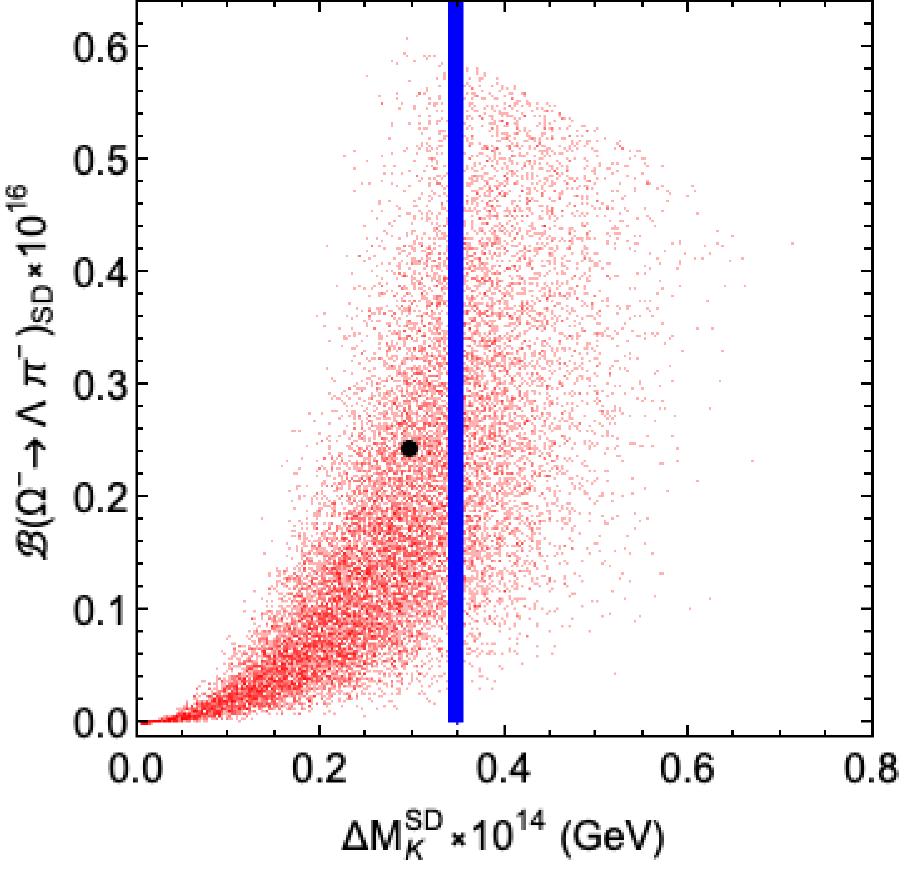}\vspace{9pt}\\ \includegraphics[width=56mm]{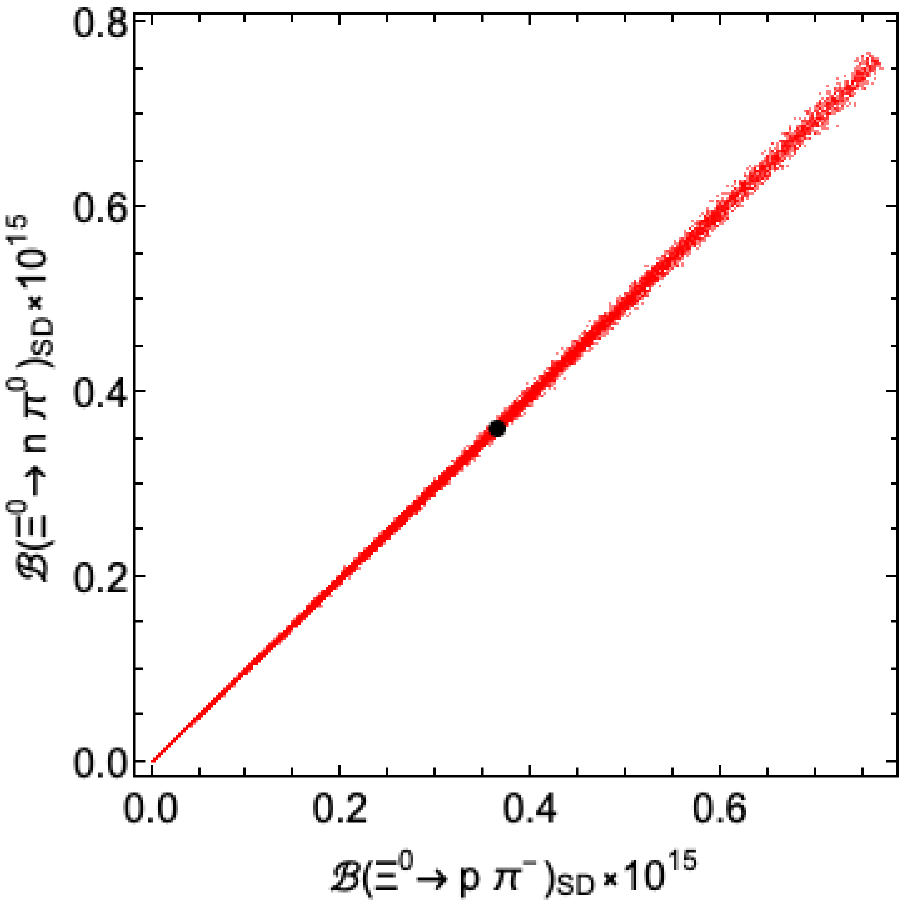} ~  \includegraphics[width=56mm]{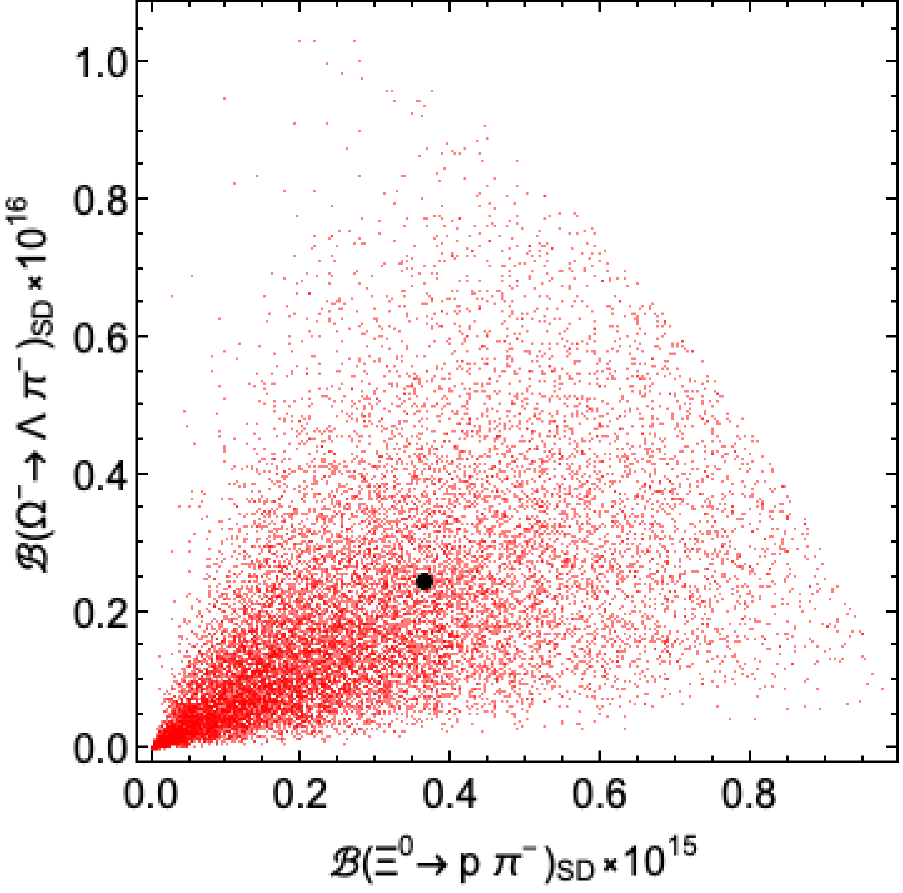} ~ \includegraphics[width=56mm]{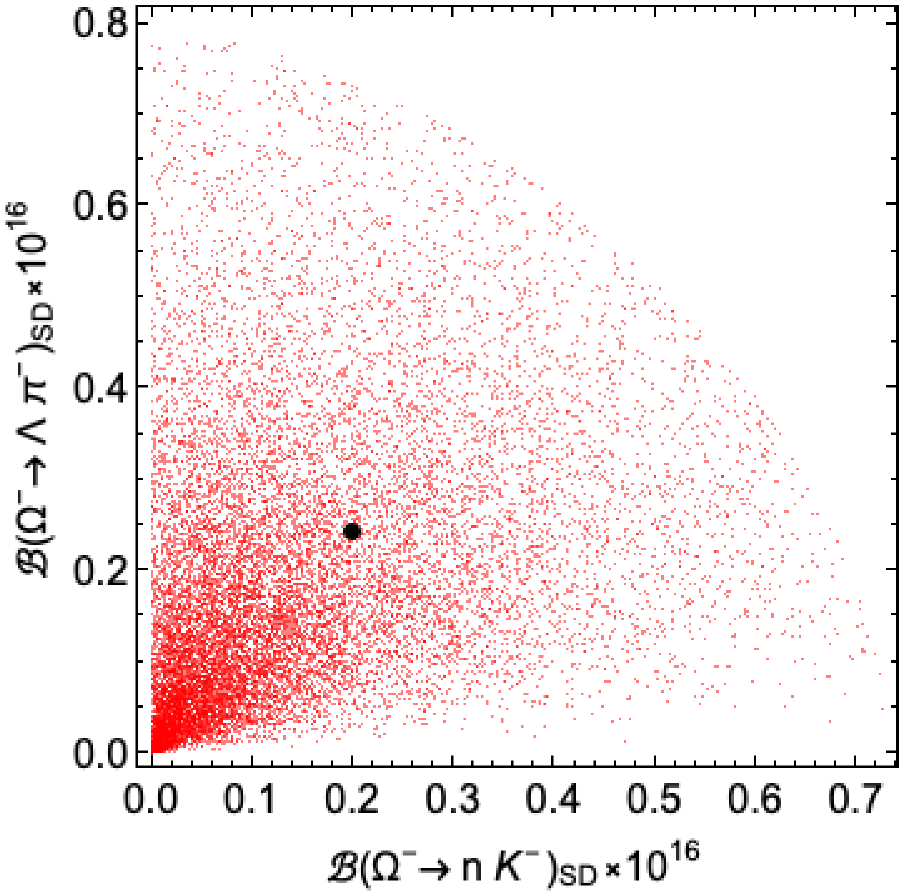}\vspace{-3pt}
\caption{Distributions (top) of ${\cal B}\big(\mathit\Xi^0\to p\pi^-\big)$ and ${\cal B}\big(\mathit\Omega^-\to\mathit\Lambda\pi^-\big)$ versus $\Delta M_K$ and (bottom) of the branching fractions of three pairs of \,$\Delta S$\,=\,2\, nonleptonic hyperon decays, all arising solely from the short-distance interactions in the SM.
The blue thick vertical lines in the top graphs indicate the experimental value, $\Delta M_K^{\rm exp}$.
The large black dots mark the central values of our estimates.\label{sdplots}}
\end{figure}

We complementarily show a number of pairwise 90\%-CL regions of quantities induced by the SM SD contributions alone in Fig.\,\ref{sdplots}, with $\hat\delta_{27}$ and $\hat\beta_{27}$ having the same sign, and of the total SM branching-fractions in Fig.\,\ref{smplots}, after applying the procedure delineated at the end of Appendix\,\,\ref{num}.
For the top (bottom) plots in Fig.\,\ref{smplots} the parameter choices are the same as those for the $\tilde{\textsc s}$ ($\tilde{\textsc p}$) column in Table\,\,\ref{t:smrates} specified in the previous paragraph.

\begin{figure}[t]
\includegraphics[width=56mm]{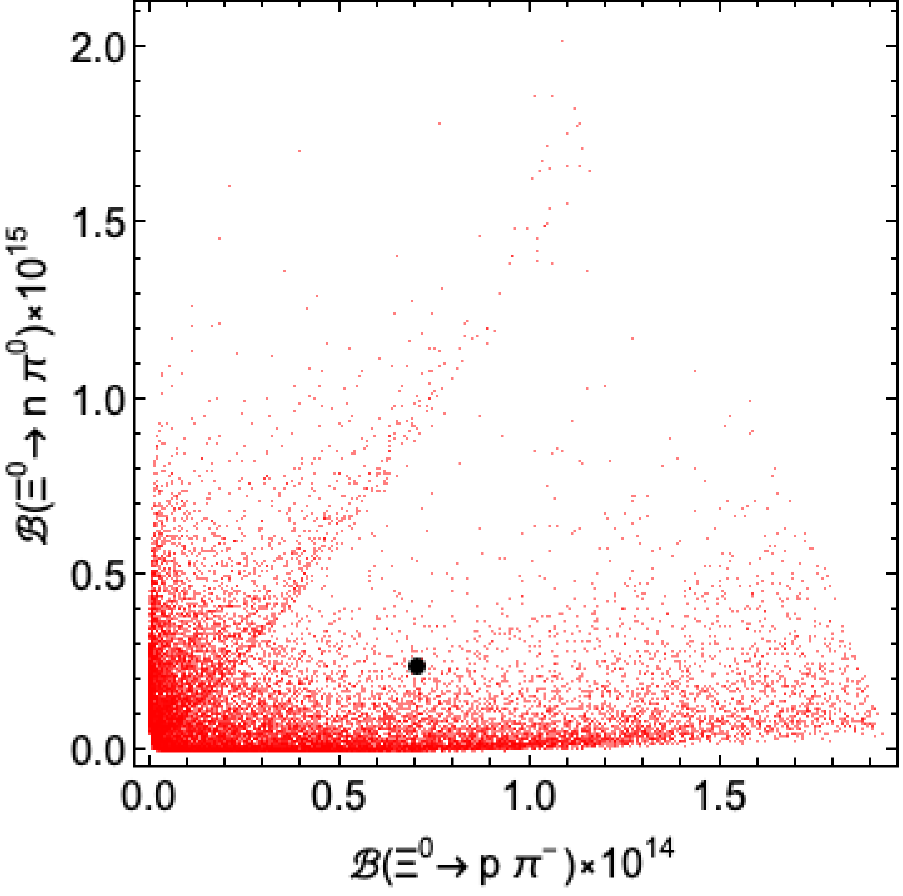} ~ \includegraphics[width=56mm]{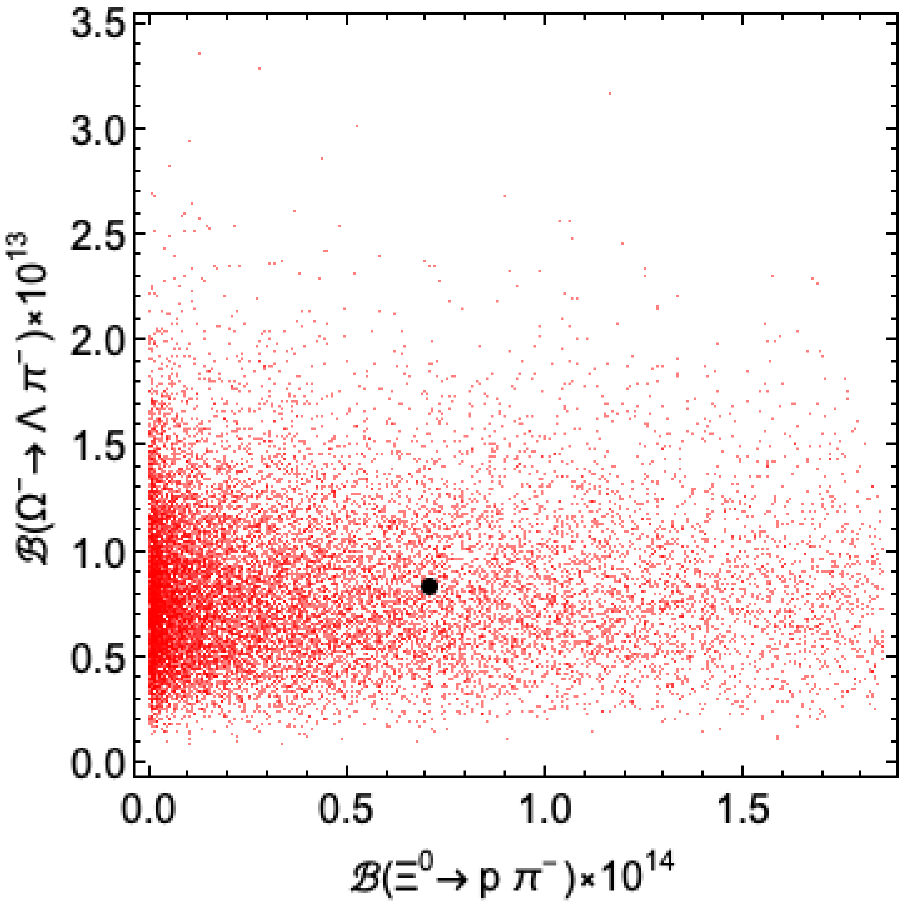} ~ \includegraphics[width=56mm]{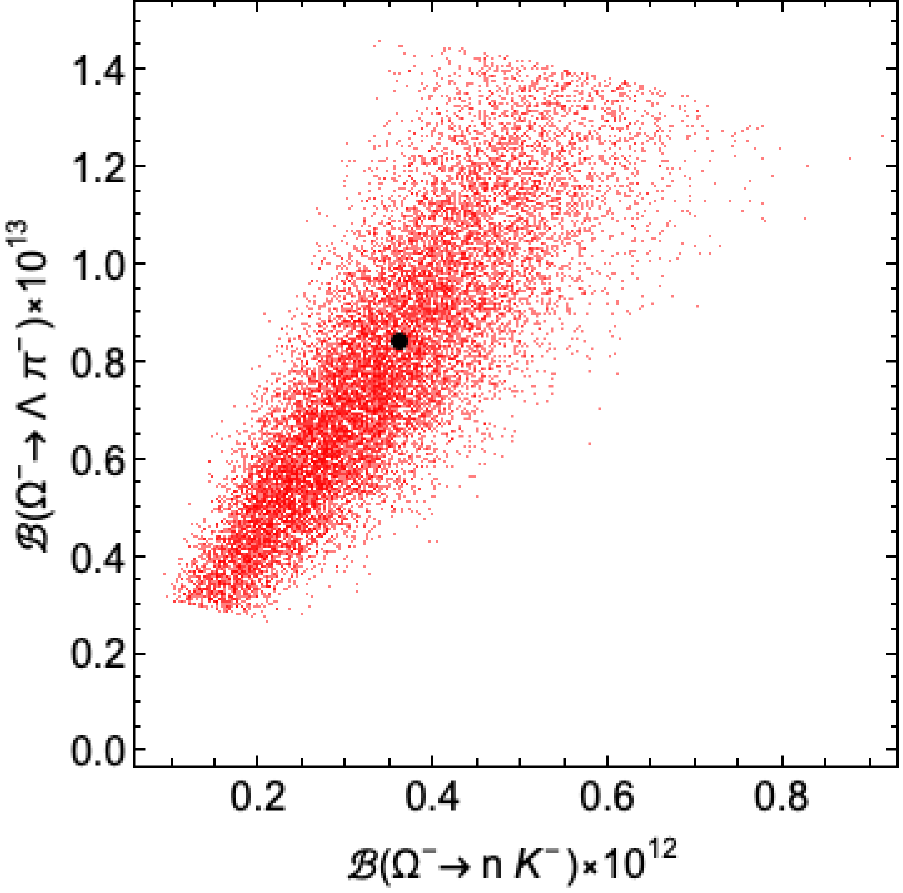}\vspace{2ex}\\\includegraphics[width=56mm]{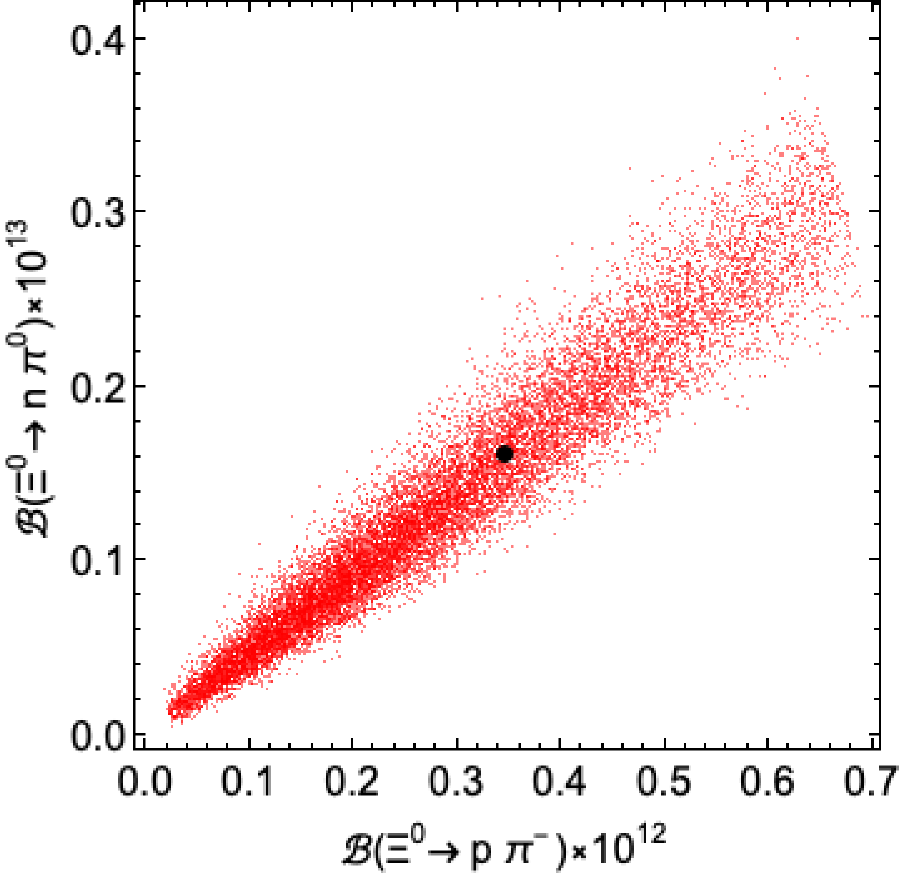} ~ \includegraphics[width=56mm]{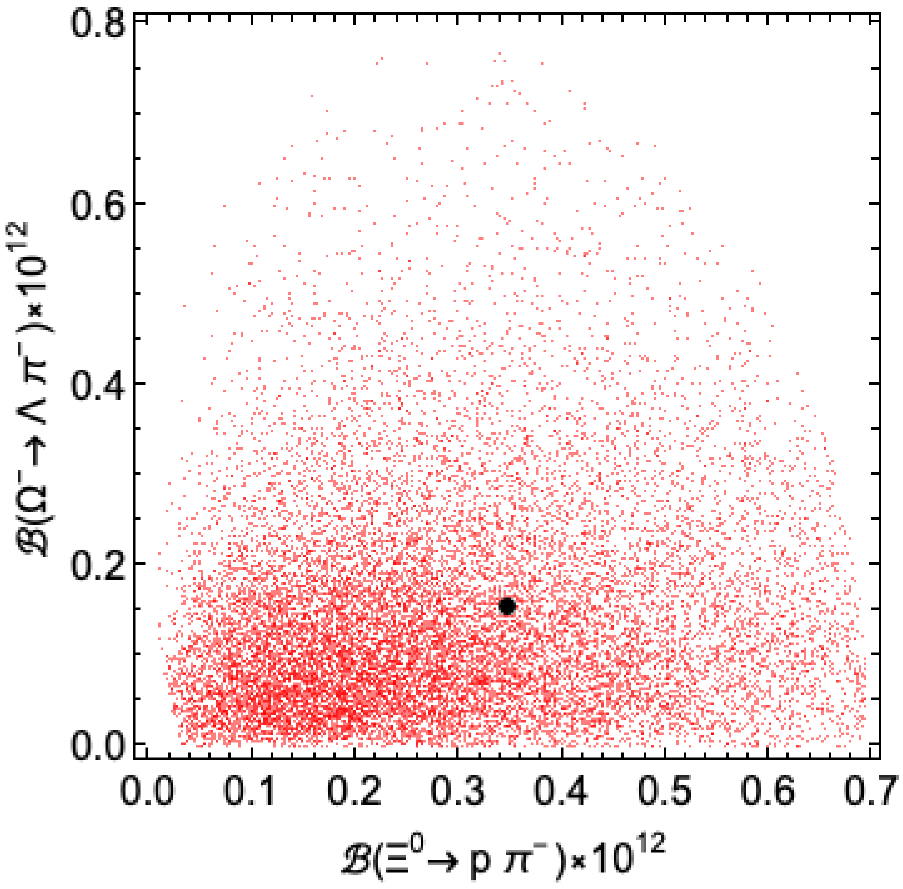} ~ \includegraphics[width=56mm]{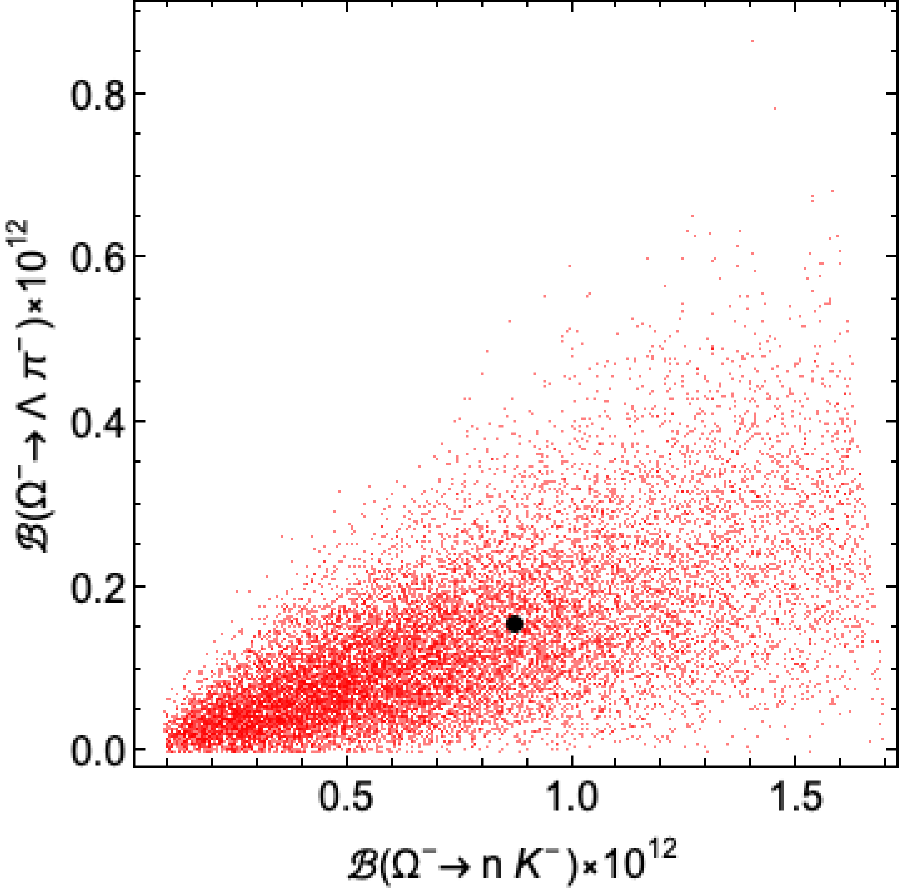}\vspace{-3pt}
\caption{Distributions of the SM branching fractions of different pairs of \,$\Delta S$\,=\,2\, nonleptonic hyperon decays from the summed SD and LD amplitudes, as explained in the text.
The large black dots mark the central values.\label{smplots}} \medskip
\end{figure}

In view of the smallness of the SM predictions in Table \ref{t:smrates}, it is unlikely that they will be testable any time soon.
On the upside, the striking dissimilarity between Eqs.\,\,(\ref{smBX2Npi}) and (\ref{smBX2Npi-v2}),
and between the corresponding entries in the third and fourth columns of Table \ref{t:smrates}, implies that future observations of \,$\mathit\Xi\to N\pi$\, with branching fractions at the level of $10^{-12}$ or below could offer extra insight for dealing with the \texttt S-wave/\texttt P-wave problem in the \,$\Delta S$\,=\,1\, nonleptonic decays of the octet hyperons.
Furthermore, given that the measured bounds on these \,$\Delta S$\,=\,2\, decays are scanty and fairly weak at the moment, the room for potential new-physics hiding in them is still substantial.

It is unfortunate that hadronic uncertainty plagues a good number of hyperon decay modes, making it difficult to tease out new-physics effects even in supposedly simpler semileptonic modes such as $\mathit\Sigma^+\to p\mu^+\mu^-$ \cite{He:2005yn,He:2018yzu,Wang:2021uzi,Geng:2021fog} or weak radiative modes \cite{Shi:2022dhw,Shi:2023kbu}.
This implies that it is essential to keep pursuing processes which in the SM are either forbidden, such as those not conserving lepton flavor/number~\cite{He:2019xxp,Li:2016tlt,Goudzovski:2022vbt} and decays into a final state containing a dark boson/fermion~\cite{Goudzovski:2022vbt,Su:2019ipw,MartinCamalich:2020dfe,Alonso-Alvarez:2021oaj}, or very rare, such as the $\Delta S$\,=\,2 ones investigated here and flavor-changing neutral-current decays with missing energy carried away by a pair of invisibles~\cite{Li:2016tlt,Geng:2021fog,Goudzovski:2022vbt,Hu:2018luj,Tandean:2019tkm,Li:2019cbk,Su:2019tjn}.
It is therefore exciting that there are ongoing and proposed quests for some of them at running facilities~\cite{Li:2016tlt,AlvesJunior:2018ldo,Goudzovski:2022vbt}.
It is also encouraging that a couple of channels that have been searched for experimentally \cite{Ang:1969hg,HyperCP:2005mvo,LHCb:2017rdd} are now under consideration by the lattice community~\cite{Erben:2022tdu}.
In addition, the aforementioned problem of \,$\Delta S$\,=\,1\, NLHD and other aspects of them continue to receive theoretical attention \cite{Wang:2019alu,Xu:2020jfr,Ivanov:2021huf,Mommers:2022dgw}.

\section{\boldmath$\Delta S$\,=\,2 nonleptonic hyperon decays from new physics\label{NP}}

The study of \,$\Delta S=2$\, processes within the SM presented in the last section serves to guide us about what can be expected with new physics (NP).
An effective theory at the weak scale required to satisfy the gauge symmetries of the SM will in general contain four-quark operators of definite chiral structure.
The \,$\Delta S$\,=\,2\, ones will then contribute to both $K^0$-$\bar K^0$ mixing and hyperon decays, and if the Wilson coefficients are constrained by the former, the latter can generally be anticipated to occur at most near SM levels.

Nevertheless, the currently huge window between the SM predictions for the hyperon modes and their empirical upper-limits invites an exploration of NP scenarios that could populate it.
It should be clear that, in order to achieve this, fine-tuning will be necessary.

We have found two ways in which NP can avoid the restriction from $K^0$-$\bar K^0$ mixing.
The first one relies on fine-tuning of model parameters that results in a cancellation among different contributions to the mixing.
This is feasible because a four-quark operator comprising purely left- or right-handed fields leads to a $K^0$-$\bar K^0$ matrix-element which is different than that of an operator consisting of chirally mixed fields.
In Sec.\,\ref{Z'} we sketch a model exemplifying how this could happen.

The second scenario was already pointed out in Ref.\,\cite{He:1997bs} and involves NP which gives rise to \,$|\Delta S|$\,=\,2\, four-quark operators that exclusively violate parity and therefore do not contribute to \,$K^0\leftrightarrow\bar K^0$\, transitions.
This also entails fine-tuning because SM gauge symmetries force any new particles to have chiral couplings to quarks at the weak scale.
Cancellations between different operators are then needed to eliminate the parity-conserving ones.
In Sec.\,\ref{LQ} we illustrate how this can be accomplished with two leptoquarks.

\subsection{\boldmath$Z'$ contributions\label{Z'}}

We entertain the possibility that there exists a spin-1 massive gauge field $Z'$ which is associated with a new Abelian gauge group U(1)$'$ and couples to SM quarks in a family-nonuniversal manner, but has negligible mixing with SM gauge bosons.
After the quark fields are rotated to the mass eigenstates, the $Z'$ gains flavor-changing interactions at tree level with generally unequal left- and right-handed couplings~\cite{Langacker:2000ju}.
Here we focus on the $dsZ'$ sector specified by the Lagrangian
\begin{align} \label{LdsZ'}
{\cal L}_{dsZ'}^{} & \,=\,
-\overline d\gamma^\beta\big(g_L^{}P_L^{}+g_R^{}P_R^{}\big)s\, Z_\beta' \,+\, {\rm H.c.} \,, &
\end{align}
with $g_L^{}$ and $g_R^{}$ being constants and \,$P_R=(1+\gamma_5)/2$.\,
We suppose that additional fermionic interactions that the $Z'$ may possess already fulfill the empirical restraints to which they are subject, but on which we do not dwell in this paper.

With the $Z'$ mass, $m_{Z'}$, assumed to be big, from Eq.\,(\ref{LdsZ'}) one can come up with tree-level $Z'$-mediated diagrams contributing to the \,$s\bar d\to\bar sd$\, reaction and described by
\begin{align} \label{Hdsds}
{\cal H}_{\Delta S=2}^{Z'} & \,=\, \frac{g_L^2 {\cal Q}_{LL}^{} + g_R^2 {\cal Q}_{RR}^{}}{2 m_{Z'}^2}
+ \frac{g_L^{}g_R^{} {\cal Q}_{LR}^{}}{m_{Z'}^2} &
\end{align}
at an energy scale \,$\mu\,\mbox{\footnotesize$\lesssim$}\,m_{Z'}$,\, with
\begin{align} \label{QRR}
{\cal Q}_{RR}^{} & \,=\, \overline d\gamma^\alpha P_R^{}s\, \overline d\gamma_\alpha^{}P_R^{}s \,, &
{\cal Q}_{LR}^{} & \,=\, \overline d\gamma^\alpha P_L^{}s\, \overline d\gamma_\alpha^{}P_R^{}s \,. &
\end{align}
To examine the effects of ${\cal H}_{\Delta S=2}^{Z'}$ on hadronic transitions, we need to take into account the QCD renormalization-group running from the $m_{Z'}$ scale down to hadronic scales.
This modifies Eq.\,(\ref{Hdsds}) into~\cite{Buras:2001ra,Buras:2012fs}
\begin{align} \label{Hdsds-loE}
{\cal H}_{\Delta S=2}^{Z'} & \,=\, \frac{\eta_{LL}^{}\, g_L^2\, {\cal Q}_{LL}^{} + \eta_{RR}^{}\, g_R^2\, {\cal Q}_{RR}^{}}{2 m_{Z'}^2} + \frac{g_L^{}g_R^{} \big( \eta_{LR}^{}\, {\cal Q}_{LR}^{} + \eta_{LR}'\, {\cal Q}_{LR}' \big)}{m_{Z'}^2} \,, &
\end{align}
where \,$\eta_{LL}^{}=\eta_{RR}^{}$\, and $\eta_{LR}^{(\prime)}$ are QCD-correction factors and \,${\cal Q}_{LR}'= \overline dP_L^{}s\,\overline dP_R^{}s$.\,

The chiral realization of ${\cal Q}_{LL}$ for hyperons is already given in Eq.\,(\ref{OLL}).
Hence, since ${\cal Q}_{RR}$ transforms like $(1_L,27_R)$ under \,${\rm SU}(3)_L\times{\rm SU}(3)_R$\, rotations and the strong interaction is invariant under a parity operation, the lowest-order chiral realization of ${\cal Q}_{RR}$ is
\begin{align} \label{ORR}
{\cal O}_{RR}^{} & \,=\, \Lambda_\chi^{}\, f_\pi^2\, \textit{\textsf t}_{kl,no}^{} \Big[ \hat\beta_{27}^{}\, \big( \xi^\dagger\overline B\xi \big)_{nk\,} \big(\xi^\dagger B\xi\big)_{ol} + \hat\delta_{27}^{}\, \xi_{nx}^\dagger \xi_{oz}^\dagger \xi_{vk}^{} \xi_{wl}^{}\, \big(\overline T_{rvw}\big){}^\eta (T_{rxz})_\eta^{} \Big] \,, &
\end{align}
For ${\cal Q}_{LR}^{(\prime)}$, which belongs to $(8_L,8_R)$ and is even under parity, the leading-order baryonic chiral realization relevant to the decays of interest is
\begin{align} \label{OLR}
{\cal O}_{LR}^{(\prime)} \,=\, \tfrac{1}{2}\Lambda_\chi^{}\, f_\pi^2\, \textit{\textsf t}_{kl,no}^{} & \Big\{ \hat\beta{}_{\scriptscriptstyle 88}^{(\prime)} \Big[ \big( \xi\overline B\xi\raisebox{0pt}{$^\dagger$} \big)_{nk}\, \big( \xi^\dagger B\xi \big)_{ol} + \big(\xi^\dagger\overline B\xi\big)_{nk}\, \big(\xi B\xi^\dagger\big)_{ol} \Big]
\nonumber \\ & \,+\, \hat\delta{}_{\scriptscriptstyle 88}^{(\prime)} \big( \xi_{nx}^{} \xi_{oz}^\dagger \xi_{vk}^\dagger \xi_{wl}^{} + \xi_{nx}^\dagger \xi_{oz}^{} \xi_{vk}^{} \xi_{wl}^\dagger \big)\, \big(\overline T_{rvw}\big){}^\eta (T_{rxz})_\eta^{} \Big\} \,, &
\end{align}
where $\hat\beta{}_{\scriptscriptstyle 88}^{(\prime)}$ and $\hat\delta{}_{\scriptscriptstyle 88}^{(\prime)}$ will be estimated shortly.
Being parity even, ${\cal O}_{LR}^{(\prime)}$ at tree level impacts only the \texttt P waves of \,$\mathit\Xi\to N\pi$\, and \,$\mathit\Omega^-\mbox{\small$\,\to\mathfrak B$}\phi,\mathit\Sigma^*\pi$.\,

It is worth commenting that the $\hat\beta{}_{\scriptscriptstyle 88}^{(\prime)}$ portion of Eq.\,(\ref{OLR}) can alternatively be expressed in terms of traces, in light of the relation  $\textit{\textsf t}_{kl,no}^{} \big(\xi\overline B\xi^\dagger\big){}_{nk}^{} \big( \xi^\dagger B\xi\big){}_{ol}^{} = {\rm Tr}\big(\hat\kappa\xi\overline B\xi^\dagger \hat\kappa\xi^\dagger B\xi\big) = {\rm Tr}\big(\hat\kappa\xi\overline B\xi^\dagger\big) {\rm Tr}\big(\hat\kappa\xi^\dagger B\xi\big)$  and the same expression but with $\xi$ and $\xi^\dagger$ interchanged.\footnote{One could construct other parity-even $(8_L,8_R)$ combinations: \,${\rm Tr}\big(\hat\kappa\mathit\Sigma\hat\kappa\mathit\Sigma^\dagger\big) {\rm Tr}\big(\overline BB\big)$,\, ${\rm Tr}\big[\overline B \big(\xi^\dagger\hat\kappa\mathit\Sigma\hat\kappa\xi^\dagger + \xi\hat\kappa\mathit\Sigma^\dagger\hat\kappa\xi\big) B \big]$,  ${\rm Tr}\big[\big(\xi^\dagger\hat\kappa\mathit\Sigma\hat\kappa\xi^\dagger + \xi\hat\kappa\mathit\Sigma^\dagger
\hat\kappa\xi\big) \overline BB\big]$,\, and \,${\rm Tr}\big( \hat\kappa\xi\overline B\xi\hat\kappa\xi^\dagger B\xi^\dagger + \hat\kappa\xi^\dagger\overline B\xi^\dagger \hat\kappa\xi B\xi \big)$.\, However, the \,$\mathit\Xi^0\to n$\, matrix-elements of the first three vanish, whereas that of the fourth is not independent from \,$\langle n|{\cal O}_{LR}^{(\prime)}|\mathit\Xi^0\rangle$\, because of the equation
\,${\rm Tr}\big[\hat\kappa\xi\overline B\xi\hat\kappa\xi^\dagger B\xi^\dagger + \hat\kappa\xi^\dagger\overline B\xi^\dagger\hat\kappa\xi B\xi + \overline B \big\{ \xi^\dagger\hat\kappa\mathit\Sigma\hat\kappa\xi^\dagger + \xi\hat\kappa\mathit\Sigma^\dagger\hat\kappa\xi, B\big\} \big] = {\rm Tr}\big( \hat\kappa\xi\overline B\xi^\dagger\hat\kappa
\xi^\dagger B\xi + \hat\kappa\xi^\dagger\overline B\xi\hat\kappa\xi B\xi^\dagger\big) + {\rm Tr}\big(\hat\kappa\mathit\Sigma\hat\kappa\mathit\Sigma^\dagger\big) {\rm Tr}\big(\overline BB\big)$.}
We further note that ${\cal Q}_{LL,RR,LR}$ and ${\cal Q}_{LR}'$ are all invariant under the $CPS$ transformation~\cite{Bernard:1985wf}, which is the ordinary $CP$ operation followed by switching the $d$ and $s$ quarks, as are their chiral realizations ${\cal O}_{LL,RR,LR}$ and ${\cal O}_{LR}'$.

With these operators, we can produce diagrams like those in Fig.\,\ref{sd-diagrams} but with the weak couplings (hollow squares) now induced by ${\cal H}_{\Delta S=2}^{Z'}$ in Eq.\,(\ref{Hdsds-loE}).
Subsequently, for \,$\mathit\Xi\to N\pi$\, we arrive at
\begin{align}
{\mathbb A}_{\mathit\Xi^0p}^{\scriptscriptstyle(Z')} & \,=\, {\mathbb A}_{\mathit\Xi^-n}^{\scriptscriptstyle(Z')} \,=\,
\frac{\textsf c_{LL}^{} - \textsf c_{RR}^{}}{2\sqrt2} \,, &
\nonumber \\
{\mathbb B}_{\mathit\Xi^0p}^{\scriptscriptstyle(Z')} & \,=\, \big(\textsf c_{LL}^{}+\textsf c_{RR}^{}+2 \textsf c_{LR}^{}+2 \textsf c_{LR}'\big)
\frac{D+F}{2\sqrt2} \bigg(\frac{m_N^{}+m_{\mathit\Xi}}{m_{\mathit\Xi}-m_N}\bigg) \,,
\nonumber \\
{\mathbb B}_{\mathit\Xi^-n}^{\scriptscriptstyle(Z')} & \,=\, \big(\textsf c_{LL}^{}+\textsf c_{RR}^{}+2 \textsf c_{LR}^{}+2 \textsf c_{LR}'\big)
\frac{D-F}{2\sqrt2} \bigg(\frac{m_N+m_{\mathit\Xi}}{m_N-m_{\mathit\Xi}}\bigg) \,,
\end{align}
where
\begin{align} \label{c's}
\textsf c_{LL(RR)}^{} & \,=\, \frac{4\pi\, \eta_{LL}^{}\, g_{L(R)}^2}{m_{Z'}^2}\, f_\pi^2\, \hat\beta_{27}^{} \,, &
\textsf c_{LR}^{(\prime)} & \,=\, \frac{4\pi\, \eta_{LR}^{(\prime)}\, g_L^{}g_R^{}}{m_{Z'}^2}\, f_\pi^2\, \hat\beta{}_{\scriptscriptstyle 88}^{(\prime)} \,. &
\end{align}
As for the $\mathit\Omega^-$ channels, we find
\begin{align} \label{z'O}
{\mathbb C}_{nK^-}^{\scriptscriptstyle(Z')} & \,=\, {\cal C}\, \frac{\textsf c_{LL}^{}+\textsf c_{RR}^{}+2 \textsf c_{LR}^{}+2 \textsf c_{LR}'}{2\sqrt2\, (m_{\mathit\Xi}-m_N)} - {\cal C}\, \frac{\tilde c_{LL}^{}+\tilde c_{RR}^{}+2 \tilde c_{LR}^{}+2\tilde c_{LR}^{\,\prime}}{6\sqrt2\, (m_\mathit\Omega-m_{\mathit\Sigma^*})} \,,
\nonumber \\ \raisebox{4ex}{}
{\mathbb C}_{\mathit\Lambda\pi^-}^{\scriptscriptstyle(Z')} & \,=\, {\cal C}\, \frac{\tilde c_{LL}^{}+\tilde c_{RR}^{}+2 \tilde c_{LR}^{}+2\tilde c_{LR}^{\,\prime}}{4\sqrt3\, (m_\mathit\Omega-m_{\mathit\Sigma^*})} \,,
\nonumber \\ \raisebox{4ex}{}
{\mathbb C}_{\mathit\Sigma^0\pi^-}^{\scriptscriptstyle(Z')} & \,=\, -{\cal C}\, \frac{\tilde c_{LL}^{}+\tilde c_{RR}^{}+2 \tilde c_{LR}^{}+2\tilde c_{LR}^{\,\prime}}{12 (m_\mathit\Omega-m_{\mathit\Sigma^*})} \,, &
\\ \raisebox{4ex}{}
\tilde{\mbox{\small$\mathbb A$}}_{\mathit\Sigma^*\pi}^{\scriptscriptstyle(Z')} & \,=\, \frac{\tilde c_{LL}^{} - \tilde c_{RR}^{}}{2\sqrt2} \,, &
\nonumber \\ \raisebox{4ex}{}
\tilde{\mbox{\small$\mathbb B$}}_{\mathit\Sigma^*\pi}^{\scriptscriptstyle(Z')} & \,=\, \frac{-\mathscr H}{6\sqrt3} \bigg(\frac{m_{\mathit\Omega}+m_{\mathit\Sigma^*}}{m_{\mathit\Omega}-m_{\mathit\Sigma^*}}\bigg) \big( \tilde c_{LL}^{}+\tilde c_{RR}^{}+2 \tilde c_{LR}^{}+2\tilde c_{LR}^{\,\prime} \big) \,,
\end{align}
where
\begin{align} \label{tc's}
\tilde c_{LL(RR)}^{} & \,=\, \frac{4\pi\, \eta_{LL}^{}\, g_{L(R)}^2}{m_{Z'}^2}\, f_\pi^2\, \hat\delta_{27}^{} \,, &
\tilde c_{LR}^{(\prime)} & \,=\, \frac{4\pi\, \eta_{LR}^{(\prime)}\, g_L^{}g_R^{}}{m_{Z'}^2}\, f_\pi^2\, \hat\delta{}_{\scriptscriptstyle 88}^{(\prime)} \,. &
\end{align}

For the coefficients in Eqs.\,\,(\ref{c's}) and (\ref{tc's}), numerically we utilize \,$\eta_{LL}^{}=0.65$, \,$\eta_{LR}^{}=0.99$,\, and \,$\eta_{LR}'=-5.08$\, evaluated at the scale \,$\mu=1\;$GeV,\, which is compatible with the fact that we implemented the techniques of chiral perturbation theory to determine the baryonic matrix elements, upon setting \,$m_{Z'}=5\;$TeV\, and employing the formulas provided by Ref.\,\cite{Buras:2001ra}.
This $m_{Z'}$ choice escapes the limitations from $Z'$ searches in hadronic final-states at colliders~\cite{ParticleDataGroup:2022pth}.
As regards $\hat\beta{}_{\scriptscriptstyle 88}^{(\prime)}$ and $\hat\delta{}_{\scriptscriptstyle 88}^{(\prime)}$, first we remark that the bag model\footnote{A textbook treatment of the bag model can be found in Ref.\,\cite{Donoghue:1992dd}.\medskip} predicts  $\hat\beta_{27}=\hat\delta_{27}=0$\, but \,$\hat\beta{}_{\scriptscriptstyle 88}=2\hat\beta{}_{\scriptscriptstyle 88}=-0.15$\, and \,$\hat\delta{}_{\scriptscriptstyle 88}=2\hat\delta{}_{\scriptscriptstyle 88}=-0.11$.\,
These and Eq.\,(\ref{hh27}), along with the expectation of naive dimensional analysis~\cite{Manohar:1983md,Georgi:1986kr} that they equal unity, then suggest that we may adopt
\begin{align} \label{88}
\hat\beta{}_{\scriptscriptstyle 88}^{} & \,=\, 2\hat\beta{}_{\scriptscriptstyle 88}' \,=\, \hat\delta{}_{\scriptscriptstyle 88}^{} \,=\, 2\hat\delta{}_{\scriptscriptstyle 88}' \,=\, 1 \mbox{~ or ~$-1$}  &
\end{align}
for our numerical work.

Before calculating the hyperon rates, we also need to pay attention to potential restrictions implied by kaon-mixing data.
This is because the interactions in Eq.\,(\ref{Hdsds-loE}) affect the neutral-kaon mass difference \,$\Delta M_K=2\,{\rm Re}\,M_{K\bar K}$\, and the $CP$-violation parameter \,$|\epsilon|\simeq|{\rm Im}\,M_{K\bar K}|/\big(\sqrt2\,\Delta M_K^{\rm exp}\big)$\, via \,$2 m_{K^0}^{} M_{K\bar K}^{\scriptscriptstyle Z'} = \langle K^0|{\cal H}_{\Delta S=2}^{Z'}|\bar K^0\rangle$.\,
Thus, the $Z'$ contribution is
\begin{align} \label{DMKZ'}
M_{K\bar K}^{\scriptscriptstyle Z'} & \,=\, \frac{\eta_{LL}^{}\, \big(g_L^2+g_R^2\big) \big\langle{\cal Q}_{LL}^{}\big\rangle + 2 g_L^{}g_R^{}\, \big(\eta_{LR}^{} \big\langle{\cal Q}_{LR}^{}\big\rangle + \eta_{LR}' \big\langle{\cal Q}_{LR}'\big\rangle\big)}{4m_{K^0}\,m_{Z'}^2} \,, &
\end{align}
where \,$\langle{\cal Q}\rangle\equiv\langle K^0|{\cal Q}|\bar K^0\rangle$.\,
Numerically \,$\langle{\cal Q}_{LL}\rangle=0.002156(34)\rm\,GeV^4$,\, $\langle{\cal Q}_{LR}\rangle=-0.0482(28)\rm\,GeV^4$,\, and \,$\langle{\cal Q}_{LR}'\rangle=0.0930(30)\rm\,GeV^4$\, computed at \,$\mu=3\;$GeV\, in Ref.\,\cite{Aebischer:2020dsw}.
In Eq.\,(\ref{DMKZ'}) we additionally use \,$\eta_{LL}^{}=0.74$, \,$\eta_{LR}^{}=0.89$,\, and \,$\eta_{LR}'=-2.07$,\, all at \,$\mu=3\;$GeV\, as well.
With these numbers, it turns out that $M_{K\bar K}^{\scriptscriptstyle Z'}$ goes to zero for certain values of \,$g_L^{}/g_R^{}$\, where one of the two couplings is small relative to the other.
In Appendix\,\,\ref{Z'model} we look at an illustrative $Z'$ model that shows in some detail how this can be realized.

\begin{figure}[h] \bigskip
\includegraphics[width=0.4\textwidth]{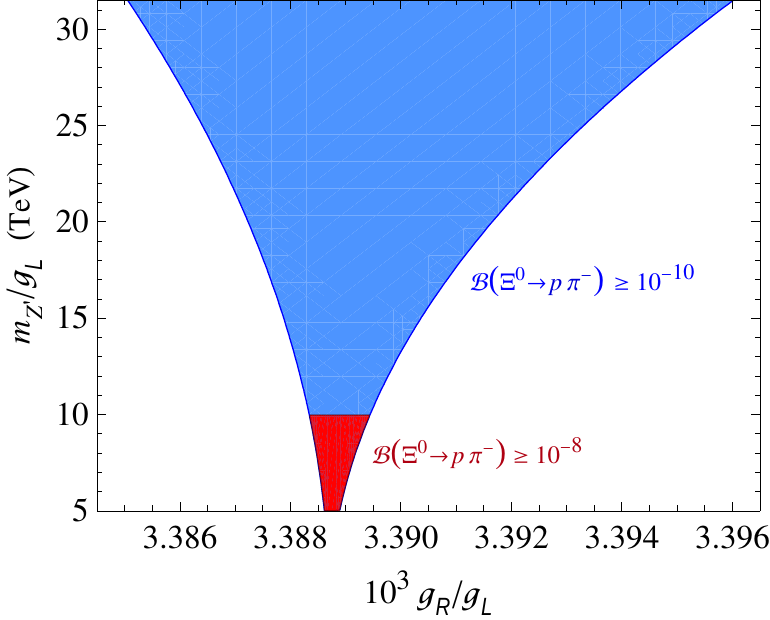} \vspace{-1ex}
\caption{Sample region of $m_{Z'}^{}/g_L^{}$ versus $g_R^{}/g_L^{}$ which can yield
\,${\cal B}(\mathit\Xi^0\to p\pi^-)_{Z'}$ between $10^{-10}$ (blue) or $10^{-8}$ (red) and $1.6\times10^{-7}$ and simultaneously satisfy the $\Delta M_K$ requirement described in the text.\label{Z'plot}}
\end{figure}

More generally, we may let $g_L^{}$ and $g_R^{}$ vary freely under the experimental requisites.
In the instance that these couplings are real, since the latest SM estimate \,$\Delta M_K^{\textsc{sm}}=5.8(2.4)\times10^{-12}\;$MeV  from lattice-QCD studies~\cite{Wang:2022lfq} is still much less precise than its measurement \,$\Delta M_K^{\rm exp}=3.484(6)\times10^{-12}\;$MeV\,\,\,\cite{ParticleDataGroup:2022pth}, we may impose \,$-1<\Delta M_K^{\scriptscriptstyle Z'}/\Delta M_K^{\rm exp}<0.5$,\, which is consistent with the two-sigma range of \,$\Delta M_K^{\rm exp}$\,$-$\,$\Delta M_K^{\textsc{sm}}$,\, but there is no constraint from $\epsilon$.
For an example of this case, we pick the first option in Eq.\,(\ref{88}) and\,\,\,$\hat\delta_{27}=-\hat\beta_{27}$,\, as well as \,$m_{Z'}^{}/g_L^{}\ge5$\,\,TeV,\, which reflects our assuming \,$|g_L^{}|\le1$\, to guarantee perturbativity, with \,$m_{Z'}=5\;$TeV.\,
This results in the allowed (blue and red) regions of $m_{Z'}^{}/g_L^{}$ versus $g_R^{}/g_L^{}$ displayed in Fig.\,\ref{Z'plot}.\footnote{By interchanging $g_L^{}$ and $g_R^{}$, one could have another allowed region, which has the same shape and size. For \,$g_{L,R}^{}<0$\, there are also two regions fulfilling the $\Delta M_K$ requirement.}
The vertical span of the red area in this figure corresponds to
\begin{align}
1.0\times10^{-8} & \,\le\, {\cal B}\big(\mathit\Xi^0\to p\pi^-\big){}_{Z'}^{} \,\le\, 1.6\times10^{-7} \,,
\nonumber \\
1.2\times10^{-8} & \,\le\, {\cal B}\big(\mathit\Xi^0\to n\pi^0\big){}_{Z'}^{} \,\le\, 1.9\times10^{-7} \,,
\nonumber \\
3.3\times10^{-9} & \,\le\, {\cal B}\big(\mathit\Xi^-\to n\pi^-\big){}_{Z'}^{} \,\le\, 5.2\times10^{-8} \,,
\\ \raisebox{4ex}{}
3.4\times10^{-9} & \,\le\, {\cal B}\big(\mathit\Omega^-\to nK^-\big){}_{Z'}^{} \,\le\, 5.4\times10^{-8} \,,
\nonumber \\
1.2\times10^{-9} & \,\le\, {\cal B}\big(\mathit\Omega^-\to\mathit\Lambda\pi^-\big){}_{Z'}^{} \,\le\, 2.0\times10^{-8} \,,
\nonumber \\
4.1\times10^{-10} & \,\le\, {\cal B}\big(\mathit\Omega^-\to\mathit\Sigma^0\pi^-\big){}_{Z'}^{} \,\le\, 6.5\times10^{-9} \,,
\nonumber \\
1.8\times10^{-9} & \,\le\, {\cal B}\big(\mathit\Omega^-\to\mathit\Sigma^{*0}\pi^-\big){}_{Z'}^{} \,\le\, 2.8\times10^{-8} \,. &
\end{align}
These are far greater than their SM counterparts in Eqs.\,(\ref{smBX2Npi-v2})-(\ref{smBO2Bphi-v2}) and might be sufficiently sizable to be within reach of LHCb~\cite{AlvesJunior:2018ldo} and BESIII~\cite{HBL} in their future quests and of the proposed Super Tau-Charm Factory~\cite{HBL}.
It should be pointed out, however, that in specific $Z'$ models the hyperon rates may be comparatively less enhanced due to various restraints on the $Z'$ couplings, such as the model discussed in Appendix\,\,\ref{Z'model}, which yields \,${\cal B}(\mathit\Xi^0\to p\pi^-)_{Z'}^{}\sim4\times10^{-10}$.\,

\subsection{Leptoquark contributions\label{LQ}}

By introducing more than one leptoquark (LQ) it is possible to generate an effective four-quark $\Delta S$\,=\,2\, interaction that is parity violating and hence eludes the kaon-mixing requirement.
The LQs of interest here, with their SM gauge-group assignments $\big({\rm SU}(3)_C,{\rm SU}(2)_L,{\rm U}(1)_Y\big)$, are  $\tilde S_1\sim(\overline3,1,4/3)$\, and \,$R_2^{}\sim(\overline3,2,7/6)$\, in the nomenclature of Ref.\,\cite{Dorsner:2016wpm}.
They can have renormalizable interactions with SM fermions according to
\begin{align} \label{tS1+R2}
{\cal L}_{\textsc{lq}}^{} & \,=\, \widetilde{\textsc y}{}_{jx\,}^{\textsc{rr}}\,
\overline{d_j^{\rm c}}\, e_x^{}\tilde S_1^{} + {\tt Y}_{jx\,}^{\texttt{LR}}\, \overline{q_j^{}}\, R_2^{}\, e_x^{} \,+\, \rm H.c. \,, &
\end{align}
where $\widetilde{\textsc y}$ and {\texttt Y} are Yukawa coupling matrices, $q_j$ and $d_j$ represent a left-handed quark doublet and right-handed down-type-quark singlet, respectively, and $e_x$ is a right-handed charged-lepton singlet.
Working in the mass basis of the down-type fermions, we rewrite Eq.\,(\ref{tS1+R2}) as
\begin{align} \label{Llq}
{\cal L}_{\textsc{lq}}^{} & \,=\, \widetilde{\textsc y}{}_{jx}^{\textsc{rr}}\,
\overline{({\texttt D}_j)^{\rm c}}P_R^{}\ell_x^{}\, \tilde S{}_1^{4/3}
+ {\tt Y}_{jx}^{\texttt{LR}} \Big( (V_{\textsc{ckm}})_{kj}^{}\, \overline{{\texttt U}_k}R_2^{5/3} + \overline{{\texttt D}_j} R_2^{2/3} \Big) P_R^{} \ell_x^{} \,+\, \rm H.c. \,, &
\end{align}
where \,$j,k,x=1,2,3$\, here denote family indices and are summed over, the superscripts of $\tilde S_1$ and $R_2$ indicate the electric charges of their components, and \,${\texttt U}_{1,2,3}=u,c,t$, \,${\texttt D}_{1,2,3}=d,s,b$,  and  $\ell_{1,2,3}=e,\mu,\tau$\, refer to the mass eigenstates.
Although these LQs could have other couplings with SM fermions or engage in scalar interactions~\cite{Dorsner:2016wpm}, for our purposes we do not entertain such possibilities, considering only the minimal ingredients already specified in ${\cal L}_{\textsc{lq}}$ above.

From Eq.\,(\ref{Llq}), with the LQs taken to be heavy, we can derive box diagrams which lead to the effective Hamiltonians
\begin{align}
{\cal H}_{\Delta S=2}^{\rm LQ} & \,=\, \frac{\big(\sum_x \widetilde{\textsc y}{}_{1x}^{\textsc{rr}*}\,
\widetilde{\textsc y}{}_{2x}^{\textsc{rr}}\big){}^2}{128\pi^2\, m_{\tilde S_1}^2}\, {\cal Q}_{RR}^{} + \frac{\big(\sum_x {\tt Y}_{1x}^{\texttt{LR}}\,{\tt Y}_{2x}^{\texttt{LR}*}\big){}^2}{128\pi^2\, m_{R_2}^2}\, {\cal Q}_{LL}^{} \,, &
\nonumber \\
{\cal H}_{\Delta C=2}^{\rm LQ} & \,=\, \frac{\big[\sum_x \big(V_{\textsc{ckm}} {\tt Y}^{\texttt{LR}}\big){}_{1x}^{} \big(V_{\textsc{ckm}}^* {\tt Y}^{\texttt{LR}*}\big){}_{2x}^{} \big]{}^2}{128\pi^2\, m_{R_2}^2}\, \overline u\gamma^\eta P_L^{}c\, \overline u\gamma_\eta^{}P_L^{}c \,,
\end{align}
where ${\cal Q}_{LL,RR}$ have been written down in Eqs.\,\,(\ref{QLL}) and (\ref{QRR}).
Evidently ${\cal L}_{\textsc{lq}}$ affects not only $\Delta M_K$ via \,$\Delta M_K^{\scriptscriptstyle\rm(LQ)}={\rm Re}\langle K^0|{\cal H}_{\Delta S=2}^{\rm LQ}|\bar K^0\rangle/m_{K^0}$\, but also its charmed-meson analog, $\Delta M_D$.

It is interesting to notice that, since $\widetilde{\textsc y}{}_{ix}^{\textsc{rr}}$ and ${\tt Y}_{ix}^{\texttt{LR}}$ besides the LQ masses are free parameters, the model parameter space contains regions in which
\,$\big(\sum_x \widetilde{\textsc y}{}_{1x}^{\textsc{rr}*}\, \widetilde{\textsc y}{}_{2x}^{\textsc{rr}}\big){}^2/m_{\tilde S_1}^2 + \big(\sum_x {\tt Y}_{1x}^{\texttt{LR}}\,{\tt Y}_{2x}^{\texttt{LR}*}\big){}^2/m_{R_2}^2$  is highly suppressed or vanishes, rendering ${\cal H}_{\Delta S=2}^{\rm LQ}$ mostly or purely parity-odd and therefore $\Delta M_K^{\scriptscriptstyle\rm(LQ)}$ also suppressed or vanishing.
In such instances the $\Delta M_K$ limitation can be evaded.\footnote{Invoking two scalar LQs to decrease certain quantities and increase others has previously been applied to other contexts \cite{Crivellin:2017zlb,Su:2019tjn}.}
In the remainder of this section, we explore this scenario and for simplicity set \,$m_{\tilde S_1}=m_{R_2}\equiv m_{\rm LQ}$\, and
\begin{align} \label{Ylq}
\widetilde{\textsc y}^{\textsc{rr}} & \,= \left(\begin{array}{ccc} 0 & 0 & y_{d\tau}^{} \vspace{1pt} \\
0 & 0 & iy_{s\tau}^{} \vspace{1pt} \\ 0 & 0 & 0 \end{array}\!\right) , &
{\tt Y}^{\texttt{LR}} & \,= \left(\begin{array}{ccc} 0 & 0 & y_{d\tau}^{} \vspace{1pt} \\
0 & 0 & y_{s\tau}^{} \vspace{1pt} \\ 0 & 0 & 0 \end{array}\!\right) , &
\end{align}
with $y_{d\tau}^{}$ and $y_{s\tau}^{}$ being real constants, ensuring that \,$\big(\mbox{\footnotesize$\sum$}_x \widetilde{\textsc y}{}_{1x}^{\textsc{rr}*}\, \widetilde{\textsc y}{}_{2x}^{\textsc{rr}}\big){}^2 + \big(\mbox{\footnotesize$\sum$}_x {\tt Y}_{1x}^{\texttt{LR}}\,{\tt Y}_{2x}^{\texttt{LR}*}\big){}^2 = 0$.\,
Hence
\begin{align} \label{VYlq}
V_{\textsc{ckm}}^{} {\tt Y}^{\tt LR} & \,= \left(\begin{array}{ccc} 0 & 0 &
V_{ud\,}^{} y_{d\tau}^{}+V_{us\,}^{} y_{s\tau}^{} \smallskip \\ 0 & 0 &
V_{cd\,}^{} y_{d\tau}^{}+V_{cs\,}^{} y_{s\tau}^{} \smallskip \\ 0 & 0 &
V_{td\,}^{} y_{d\tau}^{}+V_{ts\,}^{} y_{s\tau}^{} \end{array}\!\right) . &
\end{align}
It follows that now
\begin{align}
{\cal H}_{\Delta S=2}^{\rm LQ} & \,=\, \frac{y_{d\tau}^2\, y_{s\tau}^2}{128\pi^2\, m_{\rm LQ}^2} ({\cal Q}_{LL}-{\cal Q}_{RR}) \,, &
\\ \label{Hucuc}
{\cal H}_{\Delta C=2}^{\rm LQ} & \,=\, \frac{\big(V_{ud\,}^{}y_{d\tau}^{}+V_{us\,}^{}y_{s\tau}^{}\big)\raisebox{1pt}{$^2$}
\big(V_{cd\,}^* y_{d\tau}^{}+V_{cs\,}^* y_{s\tau}^{}\big)\raisebox{1pt}{$^2$}}{128\pi^2\, m_{\rm LQ}^2}\,
\overline u\gamma^\alpha P_L^{}c\, \overline u\gamma_\alpha^{}P_L^{}c \,.
\end{align}
Since \,${\cal Q}_{LL}-{\cal Q}_{RR}=-\overline d\gamma^\alpha s\, \overline d\gamma_\alpha^{}\gamma_5^{}s$\, is parity odd,
${\cal H}_{\Delta S=2}^{\rm LQ}$ no longer influences $K^0$-$\bar K^0$ mixing.
On the other hand, the contribution to $\Delta M_D$ is still present, but this will be avoided if one of the brackets in ${\cal H}_{\Delta C=2}^{\rm LQ}$ is zero.
Thus, we opt for \,$V_{ud\,}^{}y_{d\tau}^{}+V_{us\,}^{}y_{s\tau}^{}=0$,\, which causes \,${\cal H}_{\Delta C=2}^{\rm LQ}=0$\, and
\begin{align} \label{lqHdsds}
{\cal H}_{\Delta S=2}^{\rm LQ} & \,=\, \frac{V_{ud}^2\, y_{d\tau}^4}{128\pi^2\, m_{\rm LQ}^2 V_{us}^2} ({\cal Q}_{LL}-{\cal Q}_{RR}) &
\end{align}
at a scale \,$\mu\,\mbox{\footnotesize$\lesssim$}\,m_{\rm LQ}$.\,
Moreover, given that $V_{ud}$ and $V_{us}$ are real in the standard parametrization, $y_{d\tau}^{}$ and $y_{s\tau}^{}$ stay real as well, and with \,$V_{ud}/V_{us}=4.33$\,  from Ref.\,\cite{ParticleDataGroup:2022pth} the perturbativity condition  $|y_{d\tau,s\tau}^{}|<\sqrt{4\pi}$\, implies the requisite \,$|y_{d\tau}^{}|<0.819$.\,

It is worth remarking that in general, below the high scale $(\mu_{\rm NP})$ at which new physics is integrated out, the effects of QCD renormalization-group running on the Wilson coefficients ${\cal C}_{LL}$ and ${\cal C}_{RR}$ of ${\cal Q}_{LL}$ and ${\cal Q}_{RR}$ in the effective Hamiltonian ${\cal H}_{\rm eff}$ containing them are known to be the same~\cite{Buras:2001ra,Ciuchini:1998ix,Crivellin:2021lix}, which reflects the fact that the strong interaction conserves parity.
This means that the QCD-evolution factors, $\eta_{LL}^{}$ and $\eta_{RR}^{}$, which accompany these operators in ${\cal H}_{\rm eff}$ are also the same, \,$\eta_{LL}^{}=\eta_{RR}^{}$.\,
Then, in the case where \,${\cal C}_{LL}=-{\cal C}_{RR}$\, at $\mu_{\rm NP}$, at lower energies ${\cal H}_{\rm eff}$ is of the form \,$\eta_{LL}^{}{\cal C}_{LL}^{}{\cal Q}_{LL}^{}+\eta_{RR}^{}{\cal C}_{RR}^{}{\cal Q}_{RR}^{}=\eta_{LL}^{}{\cal C}_{LL}^{} \big({\cal Q}_{LL}^{}-{\cal Q}_{RR}^{}\big)$.\,
Accordingly, in our particular LQ scenario, Eq.\,(\ref{lqHdsds}) translates into \,$\langle K^0|{\cal H}_{\Delta S=2}^{\rm LQ}|\bar K^0\rangle=0$\, at any scale \,$\mu<m_{\rm LQ}$.\,

From the last two paragraphs and the chiral realizations of ${\cal Q}_{LL,RR}$ in Eqs.\,\,(\ref{OLL}) and (\ref{ORR}), we get the S-wave amplitude terms for \,$\mathit\Xi\to N\pi$\, and \,$\mathit\Omega^-\to\mathit\Sigma^*\pi$\,
\begin{align} \label{lqA}
{\mathbb A}_{\mathit\Xi^0p}^{\scriptscriptstyle\rm(LQ)} & \,=\, {\mathbb A}_{\mathit\Xi^-n}^{\scriptscriptstyle\rm(LQ)} \,=\,
\frac{\hat\beta{}_{27}^{}\,\eta_{LL}^{}\,f_\pi^2\,V_{ud}^2\,y_{d\tau}^4}{16\sqrt2\, \pi\, m_{\rm LQ}^2 V_{us}^2} \,, &
\nonumber \\
\tilde{\mbox{\small$\mathbb A$}}_{\mathit\Sigma^*\pi}^{\scriptscriptstyle\rm(LQ)} & \,=\, \frac{\hat\delta{}_{27}^{}\,\eta_{LL}^{}\,f_\pi^2\,V_{ud}^2\,y_{d\tau}^4}{16\sqrt3\, \pi\, m_{\rm LQ}^2 V_{us}^2} \,. &
\end{align}
In contrast, being parity odd, ${\cal H}_{\Delta S=2}^{\rm LQ}$ in Eq.\,(\ref{lqHdsds}) does not modify the \texttt P-wave parts, and consequently  ${\mathbb B}_{\mathit\Xi^0p}^{\scriptscriptstyle\rm(LQ)}={\mathbb B}_{\mathit\Xi^-n}^{\scriptscriptstyle\rm(LQ)} = {\mathbb C}_{nK^-}^{\scriptscriptstyle\rm(LQ)} = {\mathbb C}_{\mathit\Lambda\pi^-}^{\scriptscriptstyle\rm(LQ)} = {\mathbb C}_{\mathit\Sigma^0\pi^-}^{\scriptscriptstyle\rm(LQ)} = \tilde{\mbox{\small$\mathbb B$}}_{\mathit\Sigma^*\pi}^{\scriptscriptstyle\rm(LQ)} = 0$.\,

With \,$\hat\beta{}_{27}=0.076$\, as before, \,$\eta_{LL}^{}=0.68$\, for \,$m_{\rm LQ}=1$ TeV,\, and \,$|y_{d\tau}^{}|<0.8$,\, from Eq.\,(\ref{lqA}) we arrive at
\begin{align} \label{lqBXiNpi}
{\cal B}\big(\mathit\Xi^0\to p\pi^-\big)_{\rm LQ} & \,<\, 3.4\times10^{-8} \,, &
{\cal B}\big(\mathit\Xi^0\to n\pi^0\big)_{\rm LQ} & \,<\, 6.9\times10^{-8} \,, & \nonumber \\
{\cal B}\big(\mathit\Xi^-\to n\pi^-\big)_{\rm LQ} & \,<\, 2.0\times10^{-8} \,, &
{\cal B}\big(\mathit\Omega^-\to\mathit\Sigma^{*0}\pi^-\big)_{\rm LQ} & \,=\, 5.7\times10^{-9} \,,
\end{align}
the upper values exceeding the corresponding SM predictions in Eqs.\,(\ref{smBX2Npi-v2})-(\ref{smBO2Bphi-v2}) and Table\,\,\ref{t:smrates} by five orders of magnitude or more.
Some of these enhanced results might soon be probed by LHCb~\cite{AlvesJunior:2018ldo} and BESIII~\cite{HBL}.

Finally, we comment that although the LQs considered here influence various other low-energy processes, such as \,$s\to d\gamma,dg$\, and the anomalous magnetic moment of the $\tau$ lepton, we have checked that the effects are not significant with the parameter choices we made.
These include the special textures of the Yukawa matrices in Eq.\,(\ref{Ylq}) which also help the LQs evade the constraints from collider quests~\cite{ParticleDataGroup:2022pth}.

\section{Conclusions\label{concl}}

We have explored the \,$\Delta S$\,=\,2\, nonleptonic decays of the lowest-mass hyperons within and beyond the SM.
Concentrating on two-body channels, we first updated the SM predictions for \,$\mathit\Xi\to N\pi$\, and subsequently addressed those for \,$\mathit\Omega^-\to nK^-,\mathit\Lambda\pi^-,\mathit\Sigma\pi,\mathit\Sigma^*\pi$.\,
Furthermore, we investigated the impact on these processes of long-distance diagrams involving two couplings from the \,$\Delta S$\,=\,1  Lagrangian in the SM.
The LD contributions turned out to be much bigger than the SD ones on the whole, but can raise the branching fractions of the majority of these decay modes merely to the $10^{-12}$ level, making the SM predictions unlikely to be tested in the near future.
Beyond the SM, new physics may bring about substantial amplifications, although restrictions from kaon mixing play a consequential role.
We showed that a $Z'$ boson possessing family-nonuniversal interactions with quarks can give rise to rates of the \,$\Delta S$\,=\,2\, hyperon transitions which greatly surpass the SM expectations and a few of which could be within reach of BESIII and LHCb.
We also demonstrated that a model with two leptoquarks can achieve similar outcomes.
Although these two cases are very distinct in their details, both require some degree of fine-tuning to make the hyperon modes potentially observable not too long from now.

\acknowledgments

We thank Ulrik Egede and Hai-Bo Li for information on experimental matters.
JT thanks the Tsung-Dao Lee Institute, Shanghai Jiao Tong University, and the Hangzhou Institute for Advanced Study, University of Chinese Academy of Sciences, for their hospitality during the completion of this paper.
This work was supported in part by the Fundamental Research Funds for the Central Universities and in part by the Australian Government through the Australian Research Council Discovery Project DP200101470.
XGH was supported in part by the NSFC (Grant Nos. 11735010, 11975149, and 12090064) and in part by the MOST (Grant No. MOST 109-2112-M-002-017-MY3).

\appendix

\section{Numerical input\label{num}}

For our SM estimates, we use \,$f_\pi^{}=(92.07\pm0.85)$\,MeV\, and \,$m_c^{}=(1.27\pm0.02)$\,GeV,\, as well as  \,$G_{\rm F}=1.1663788(6)\times10^{-5}\rm\,GeV^{-2}$,\, from the Particle Data Group \cite{ParticleDataGroup:2022pth}, which also supplies the CKM factors \,$V_{ud}^*V_{us}^{}=0.21923(62)$\, and \,$|V_{cd}V_{cs}|=0.21890(61)$\, and the values of hadron masses and hyperon lifetimes.
For other parameters relevant to the SD amplitudes, we employ
\begin{align} \label{DFCH}
\beta_{27}^{}  & \,=\, 0.076\pm0.015 \,, & \delta_{27}^{} & \,=\, \pm(0.076\pm0.015) \,, &
\eta_{cc}^{} & \,=\, 1.87\pm 0.76 \,, ~~~ \nonumber \\
D & \,=\, 0.81\pm 0.01 \,, & F & \,=\, 0.47\pm 0.01 \,, & \nonumber \\
\cal C & \,=\, -1.7\pm 0.3 \,, & \mathscr H & \,=\, -2.6\pm0.5 \,,
\end{align}
where $\eta_{cc}^{}$ was computed in Ref.\,\cite{Brod:2011ty} and $D$ and $F$ ($\cal C$) were inferred at leading order from the data~\cite{ParticleDataGroup:2022pth} on semileptonic octet-baryon decays (strong decays \,$T\mbox{\small$\,\to\mathfrak B$}\phi$\, of the decuplet baryons), but we have adopted \,${\mathscr H}=3\,{\cal C}/2$\, from the nonrelativistic quark model~\cite{Jenkins:1991es} because it also predicts  $2D=3F$\, and \,${\cal C}=-2D$\, which are reasonably fulfilled by Eq.\,(\ref{DFCH}) and an empirical tree-level value of $\mathscr H$ is not yet available.

For $h_D$, $h_F$, and $h_C$, which enter the LD amplitudes, we use one of two sets of numbers resulting from fitting to either the \texttt S waves or the \texttt P waves of $\Delta S$\,=\,1 nonleptonic octet-hyperon decays and also to the \texttt P waves of \,$\Delta S$\,=\,1\, $\mathit\Omega^-\mbox{\small$\,\to\mathfrak B$}\phi$.\,
The central values and variance-covariance matrices for these two cases are, respectively,
\begin{align}
(h_D,h_F,h_C) & \,=\, (-1.69,3.96,3.75)\times 10^{-8} \,, \nonumber \\
\sigma & \,=\, \left( \begin{array}{ccc}
 0.189 & 0.023  & 0.005   \\
  0.023& 0.050  & -0.006  \\
 0.005 & -0.006  &   0.121 \end{array} \right) \times 10^{-16} \,, \label{sfit}
\\ \raisebox{4ex}{}
(h_D,h_F,h_C) & \,=\, (-4.33,5.67,3.40)\times 10^{-8} \,, \nonumber \\
\sigma & \,=\, \left( \begin{array}{ccc}
 0.610 & -0.756  & 0.130   \\
  -0.756 & 0.952   & -0.164  \\
 0.130 & -0.164  &   1.256 \end{array} \right)\times 10^{-16} \,, & \label{pfit}
\end{align}
obtained from weighted least-squares fits, after the experimental errors in the weights were increased to 20\% to account for those errors being far smaller than the expected theoretical uncertainty~\cite{Jenkins:1991bt}.

To estimate the ranges in Table\,\,\ref{t:smrates}, we first assume that the errors in the input parameters listed in the previous two paragraphs are Gaussian.
We then combine these errors by generating a large sample of $k$ observable values and extracting from them the confidence-level regions.
The 90\%-CL interval range is determined by dropping the lowest and highest 5\% of the simulated values.

For the graphs in Figs.\,\,\ref{sdplots} and \ref{smplots}, we define a distance between each generated pair $(o_{i},o_{j})_k$ and their mean  $(\bar o_i,\bar o_j)$ as
\begin{align}
\tilde d(k) & \,=~ \raisebox{3pt}{\footnotesize$\displaystyle\sum_{i,j}$}\, (o_{i}-\bar o_i)\, \tilde\Sigma{}^{-1}_{ij}\, (o_{j}-\bar o_j) \,,
\end{align}
where $\tilde\Sigma_{12}$ is the numerically estimated variance-covariance matrix for the pair.
Then we select the 90\% of points closest to their mean.

\section{Rates of \boldmath$\Omega^-$ decays}

The amplitudes for \,$\mathit\Omega^-\to\mathfrak B\phi$\, and \,$\mathit\Omega^-\to\mathit\Sigma^{*0}\pi^-$\, are
\begin{align} \label{Oamps}
i{\cal M}_{\mathit\Omega^-\to\mathfrak B\phi}^{} & \,=\, {\mathbb C}_{\mathfrak B\phi}^{}\,\bar u_{\mathfrak B}^{}\,u_{\mathit\Omega}^\alpha\, \tilde p_\alpha^{} \,, &
i{\cal M}_{\mathit\Omega^-\to\mathit\Sigma^{*0}\pi^-}^{} & \,=\, \bar u_{\mathit\Sigma^*}^\alpha \big( \tilde{\mbox{\small$\mathbb A$}}_{\mathit\Sigma^*\pi}^{} - \gamma_5^{}\, \tilde{\mbox{\small$\mathbb B$}}_{\mathit\Sigma^*\pi}^{} \big) u_{\mathit\Omega,\alpha}^{} \,, ~~~
\end{align}
where ${\mathbb C}_{\mathfrak B\phi}$, $\tilde{\texttt A}_{\mathit\Sigma^*\pi}$, and $\tilde{\texttt B}_{\mathit\Sigma^*\pi}$ are constants, $\tilde p$ stands for the momentum of $\phi$, and the \texttt D-wave term in ${\cal M}_{\mathit\Omega^-\to\mathfrak B\phi}$ and the \texttt D-wave and \texttt F-wave ones in ${\cal M}_{\mathit\Omega^-\to\mathit\Sigma^{*0}\pi^-}$ have been neglected.
To calculate the corresponding rates, we need the sum over polarizations, $\varsigma$, of a spin-3/2 particle of momentum \texttt k and mass \texttt m given by\footnote{This can be found in, {\it e.g.}, Ref.\,\cite{Christensen:2013aua}.}
\begin{align} \label{pols}
\raisebox{2pt}{\scriptsize$\displaystyle\sum_{\displaystyle\varsigma=-3/2}^{\displaystyle3/2}$}\, u^\mu(\texttt k,\varsigma)\, \bar u^\nu(\texttt k,\varsigma)
& \,=\, (\slashed{\texttt k} + \texttt m) \bigg( \frac{\gamma_\rho\gamma_\omega}{3}\, {\mathcal G}^{\rho\mu} {\mathcal G}^{\omega\nu} - {\mathcal G}^{\mu\nu} \bigg) \,, &
{\mathcal G}^{\mu\nu} & \,=\, g^{\mu\nu} - \frac{\texttt k^\mu\texttt k^\nu}{\texttt m^2} \,.
\end{align}
After averaging (summing) the absolute squares of the amplitudes over the initial (final) spins, we arrive at
\begin{align} \label{GO2Bphi} &
\Gamma_{\mathit\Omega^-\to\mathfrak B\phi}^{} \,=\, \frac{{\mathcal E}+m_{\mathfrak B}^{}}{12\pi m_{\mathit\Omega}^{}}\, |\mbox{\footnotesize\boldmath$\mathcal P$}|^3\, \big|{\mathbb C}_{\mathfrak B\phi}\big|\raisebox{2pt}{$^2$} \,, &
\\ \label{GO2Sspi}
\Gamma_{\mathit\Omega^-\to\mathit\Sigma^{*0}\pi^-}^{} & \,=\, \frac{|\mbox{\footnotesize\boldmath$\mathcal P$}|}{72\pi\, m_{\mathit\Omega}^2} \Bigg[ \Bigg( \frac{\tilde\mu_+^6 + \tilde\mu_-^2\, \tilde\mu_+^4}{4\, m_{\mathit\Omega}^2\, m_{\mathit\Sigma^{*0}}^2} + 5\, \tilde\mu_+^2 \Bigg) \big|\tilde{\mbox{\small$\mathbb A$}}_{\mathit\Sigma^*\pi}^{}\big|\raisebox{2pt}{$^2$} + \Bigg( \frac{\tilde\mu_-^6 + \tilde\mu_-^4\, \tilde\mu_+^2}{4\, m_{\mathit\Omega}^2\, m_{\mathit\Sigma^{*0}}^2} + 5\, \tilde\mu_-^2 \Bigg) \big|\tilde{\mbox{\small$\mathbb B$}}_{\mathit\Sigma^*\pi}^{}\big|\raisebox{2pt}{$^2$} \Bigg] \,,
\end{align}
where \,$\tilde\mu_\pm^2=(m_{\mathit\Omega}\pm m_{\mathit\Sigma^{*0}})^2-m_{\pi^-}^2$\, and $\mathcal E$ ({\footnotesize\boldmath$\mathcal P$}) is the energy (three-momentum) of the daughter baryon in the $\mathit\Omega^-$ rest-frame.

\section{Simple \boldmath$Z'$ possibility\label{Z'model}}

For a particular example of the $Z'$ scenario considered in Sec.\,\ref{Z'}, we suppose that under the U(1)$'$ gauge group the left- and right-handed quarks in the first (second) family carry charge  $\acute{\textsc q}=1\,(-1)$\, whereas the other SM fermions are singlets.
It is straightforward to see that with these charge assignments the model is free of gauge anomalies.
Accordingly, with the covariant derivative of fermion {\texttt f} having the form \,${\cal D}_\alpha^{}{\texttt f} \supset \big(\partial_\alpha^{}+i\acute g\,\acute{\textsc q}_{\texttt f} Z_\alpha'\big){\texttt f}$,\, the $Z'$ interactions with the quarks are described by
\begin{align} \label{LqZ'}
-{\cal L}_{qZ'}^{} & \,=\, \acute g \big( \overline{u_L'}\gamma^\eta u_L' - \overline{c_L'}\gamma^\eta c_L' + \overline{d_L'}\gamma^\eta d_L' - \overline{s_L'}\gamma^\eta s_L' \,+\, \mbox{\footnotesize$(L\to R)$} \big) Z_\eta'
\nonumber \\ & \,=\,
\acute g \big[ \overline{{\texttt U}_L^{}} \gamma^\eta\, {\mathscr V}_L^{u\dagger}\, {\rm diag}(1,-1,0)\,{\mathscr V}_L^u {\texttt U}_L^{} + \overline{{\texttt D}_L^{}} \gamma^\eta\, {\mathscr V}_L^{d\dagger}\, {\rm diag}(1,-1,0)\,{\mathscr V}_L^d {\texttt D}_L^{} \,+\, \mbox{\footnotesize$(L\to R)$} \big] Z_\eta' \,, &
\end{align}
where $\acute g$ denotes the U(1)$'$ gauge coupling constant, the primed quark fields are in the flavor basis,  {\texttt U} and {\texttt D} represent column matrices with elements \,$({\texttt U}_1,{\texttt U}_2,{\texttt U}_3)=(u,c,t)$\, and \,$({\texttt D}_1,{\texttt D}_2,{\texttt D}_3)=(d,s,b)$  in the mass basis, and ${\mathscr V}_{L,R}^u$ and ${\mathscr V}_{L,R}^d$ are 3$\times$3 unitary matrices which connect the fields in the two bases and also diagonalize the quark mass matrices $M_u$ and $M_d$ via \,${\rm diag}(m_u,m_c,m_t)={\mathscr V}_L^{u\dagger}M_u^{}{\mathscr V}_R^u$  and \,${\rm diag}(m_d,m_s,m_b)={\mathscr V}_L^{d\dagger}M_d^{}{\mathscr V}_R^d$.\,

Since ${\mathscr V}_L^{u,d}$ are linked to the CKM matrix by \,${\mathscr V}_L^d={\mathscr V}_L^u V_{\textsc{ckm}}^{}$,\, the expression for ${\mathscr V}_L^d$ is fixed once ${\mathscr V}_L^u$ has been specified and vice versa, but this does not apply to ${\mathscr V}_R^{u,d}$ and there is freedom to pick their elements.
This is because $M_{u,d}$ are arbitrary as long as they satisfy the abovementioned diagonalization equations and can be arranged to have the desired textures by introducing the appropriate Higgs sector.
To suppress other effects of the new Higgs particles, including flavor-changing neutral currents which might be associated with them, they are assumed to be sufficiently heavy.

Thus, for our purposes, we can choose
\begin{align} \label{V}
{\mathscr V}_{L,R}^u & \,= \left(\!\begin{array}{ccc} \cos\theta_{L,R}^u & \sin\theta_{L,R}^u & 0 \vspace{3pt} \\ -{\rm sin}\,\theta_{L,R}^u & ~\cos\theta_{L,R}^u~ & 0 \vspace{1pt} \\ 0 & 0 & 1 \end{array}\right) , &
{\mathscr V}_R^d & \,= \left(\!\begin{array}{ccc} \cos\theta_R^d & e^{i\omega}\,\sin\theta_R^d & 0 \vspace{3pt} \\ -e^{-i\omega}\, {\rm sin}\,\theta_R^d & ~\cos\theta_R^d~ & 0 \vspace{1pt} \\ 0 & 0 & 1 \end{array}\right) , &
\end{align}
with which Eq.\,(\ref{LqZ'}) becomes
\begin{align}
-{\cal L}_{qZ'}^{} & \,=\, \acute g\, \big\{ \overline u\gamma^\eta\big({\mathscr C}_L^u P_L^{}+{\mathscr C}_R^u P_R^{}\big)u - \overline c\gamma^\eta\big({\mathscr C}_L^u P_L^{}+{\mathscr C}_R^u P_R^{}\big)c + \big[\overline u\gamma^\eta\big({\mathscr S}_L^u P_L^{}+{\mathscr S}_R^u P_R^{}\big)c \,+\, {\rm H.c.} \big] \big\} Z_\eta'
\nonumber \\ & ~~~ +\,
\acute g\, \big\{ \big[ \big(|V_{ud}|^2-|V_{cd}|^2\big) {\mathscr C}_L^u + 2\, {\rm Re}\big(V_{ud}^*V_{cd}^{}\big) {\mathscr S}_L^u \big] \overline{d_L^{}}\gamma^\eta d_L^{} + {\mathscr C}_R^d\, \overline{d_R^{}}\gamma^\eta d_R^{} \big\} Z_\eta'
\nonumber \\ & ~~~ +\,
\acute g\, \big\{ \big[ \big(|V_{us}|^2-|V_{cs}|^2\big) {\mathscr C}_L^u + 2\, {\rm Re}\big(V_{us}^*V_{cs}^{}\big) {\mathscr S}_L^u \big] \overline{s_L^{}}\gamma^\eta s_L^{} - {\mathscr C}_R^d\, \overline{s_R^{}}\gamma^\eta s_R^{} \big\} Z_\eta'
\nonumber \\ & ~~~ +\,
\acute g\, \big[ \big(|V_{ub}|^2-|V_{cb}|^2\big) {\mathscr C}_L^u + 2\, {\rm Re}\big(V_{ub}^*V_{cb}^{}\big) {\mathscr S}_L^u \big] \overline{b_L^{}}\gamma^\eta b_L^{}\, Z_\eta'
\nonumber \\ & ~~~ +\,
\acute g\, \big\{ \big[ \big(V_{ud}^*V_{us}^{}-V_{cd}^*V_{cs}^{}\big) {\mathscr C}_L^u + \big(V_{ud}^*V_{cs}^{}
+ V_{cd}^*V_{us}^{}\big) {\mathscr S}_L^u \big] \overline{d_L^{}}\gamma^\eta s_L^{} + e^{i\omega}\, {\mathscr S}_R^d\, \overline{d_R^{}}\gamma^\eta s_R^{} \,+\, {\rm H.c.} \big\}Z_\eta'
\nonumber \\ & ~~~ +\,
\acute g\, \big\{ \big[ \big(V_{ud}^*V_{ub}^{}-V_{cd}^*V_{cb}^{}\big) {\mathscr C}_L^u + \big(V_{ud}^*V_{cb}^{}
+ V_{cd}^*V_{ub}^{}\big) {\mathscr S}_L^u \big] \overline{d_L^{}}\gamma^\eta b_L^{} \,+\, {\rm H.c.} \big\}Z_\eta'
\nonumber \\ & ~~~ +\,
\acute g\, \big\{ \big[ \big(V_{us}^*V_{ub}^{}-V_{cs}^*V_{cb}^{}\big) {\mathscr C}_L^u + \big(V_{us}^*V_{cb}^{}
+ V_{cs}^*V_{ub}^{}\big) {\mathscr S}_L^u \big] \overline{s_L^{}}\gamma^\eta b_L^{} \,+\, {\rm H.c.} \big\}Z_\eta' \,, &
\end{align}
where the $\theta$s and $\omega$ are real quantities, \,${\mathscr C}_{\texttt X}^{\texttt f}={\rm cos}\big(2\theta_{\texttt X}^{\texttt f}\big)$,\, and \,${\mathscr S}_{\texttt X}^{\texttt f}={\rm sin}\big(2\theta_{\texttt X}^{\texttt f}\big)$.\,
Taking $\theta_L^u$ and $\theta_R^u$ to be tiny or vanishing then leads to
\begin{align} \label{qqZ'}
-{\cal L}_{qZ'}^{} & \,\simeq\, \acute g\, \big[ \overline u\gamma^\eta u - \overline c\gamma^\eta c + \big(|V_{ud}|^2-|V_{cd}|^2\big) \overline{d_L^{}}\gamma^\eta d_L^{} + {\mathscr C}_R^d\, \overline{d_R^{}}\gamma^\eta d_R^{} \big] Z_\eta'
\nonumber \\ & ~~~ +\,
\acute g\, \big[ \big(|V_{us}|^2-|V_{cs}|^2\big) \overline{s_L^{}}\gamma^\eta s_L^{} - {\mathscr C}_R^d\, \overline{s_R^{}}\gamma^\eta s_R^{} + \big(|V_{ub}|^2-|V_{cb}|^2\big)\, \overline{b_L^{}}\gamma^\eta b_L^{} \big] Z_\eta'
\nonumber \\ & ~~~ +\,
\acute g\, \big[ \big(V_{ud}^*V_{us}^{}-V_{cd}^*V_{cs}^{}\big) \overline{d_L^{}}\gamma^\eta s_L^{} + e^{i\omega}\, {\mathscr S}_R^d\, \overline{d_R^{}}\gamma^\eta s_R^{} \,+\, {\rm H.c.} \big] Z_\eta'
\nonumber \\ & ~~~ +\,
\acute g\, \big[ \big(V_{ud}^*V_{ub}^{}-V_{cd}^*V_{cb}^{}\big) \overline{d_L^{}}\gamma^\eta b_L^{} + \big(V_{us}^*V_{ub}^{}-V_{cs}^*V_{cb}^{}\big) \overline{s_L^{}}\gamma^\eta b_L^{} \,+\, {\rm H.c.} \big] Z_\eta' \,, &
\end{align}
where the $ucZ'$ part has dropped out, avoiding the limitation from $D^0$-$\bar D^0$ mixing.
Comparing the $dsZ'$ portion of Eq.\,(\ref{qqZ'}) with Eq.\,(\ref{LdsZ'}), we identify \,$g_L^{}=\acute g\, \big(V_{ud}^*V_{us}^{}-V_{cd}^*V_{cs}^{}\big)$\, and \,$g_R^{}=e^{i\omega} \acute g\,{\mathscr S}_R^d$.\,
Selecting \,$\omega={\rm Arg}\big(V_{ud}^*V_{us}^{}-V_{cd}^*V_{cs}^{}\big)$\, and a suitable $\theta_R^d$, we can then acquire the special $g_L^{}/g_R^{}$ ratio which renders $M_{K\bar K}^{\scriptscriptstyle Z'}$ in Eq.\,(\ref{DMKZ'}) vanishing.

It is interesting to point out that, after the CKM parameters from Ref.\,\cite{ParticleDataGroup:2022pth} are incorporated, the terms \,${\cal L}_{qZ'}\supset -\acute g\, (0.011-0.003 i)\, \overline{d_L}\slashed Z{}' b_L + \acute g\, (0.040+0.0008 i)\, \overline{s_L}\slashed Z{}'b_L + {\rm H.c.}$\, can be shown to elude $B_{d,s}$-$\bar B_{d,s}$ mixing constraints if \,$\acute g/m_{Z'}\,\mbox{\footnotesize$\lesssim$}\,0.1$/TeV,\,  as new-physics effects of order \,{\footnotesize$\sim$\,}10\% in the mass differences $\Delta M_{d,s}$ are still permitted~\cite{DeBruyn:2022zhw}.
Moreover, although a flavor-changing coupling and a~flavor-diagonal one from Eq.\,(\ref{qqZ'}) can translate into operators contributing to four-quark penguin interactions~\cite{Buchalla:1995vs}, the impact can be demonstrated to be weaker than that of the SM by at least an order of  magnitude if \,$\acute g/m_{Z'}\,\mbox{\footnotesize$\lesssim$}\,0.1$/TeV.\,
In addition, the flavor-conserving couplings in Eq.\,(\ref{qqZ'}) can escape the restraints from $Z'$ searches in hadronic final-states at colliders provided that the $Z'$ mass is around 5\,\,TeV or more~\cite{ParticleDataGroup:2022pth}.

Lastly, from Eq.\,(\ref{qqZ'}) one can derive long-distance contributions to \,$\Delta S$\,=\,2\, transitions involving two \,$\Delta S$\,=\,1\, $Z'$-mediated couplings or one of them and one \,$\Delta S$\,=\,1\, coupling from the SM.
One can deduce from the preceding two paragraphs, however, that such LD effects are unimportant relative to the SD interactions in Eq.\,(\ref{Hdsds-loE}).

\end{document}